\RequirePackage{ifpdf}
\documentclass{JHEP3}

\usepackage{arydshln}

\usepackage{amsfonts,amssymb,amsmath,bm}
\usepackage{slashed}
\usepackage{graphicx}

\ifpdf
{}
\else
\fi

\usepackage{epstopdf}

\newcommand{\beq}{\begin{equation}}
\newcommand{\eeq}{\end{equation}}
\newcommand{\beqq}{\begin{equation*}}
\newcommand{\eeqq}{\end{equation*}}
\newcommand\beqa{\begin{eqnarray}}
\newcommand\eeqa{\end{eqnarray}}
\newcommand\beqaa{\begin{eqnarray*}}
\newcommand\eeqaa{\end{eqnarray*}}
\newcommand\bea{\begin{array}}
\newcommand\eea{\end{array}}
\newcommand\beaa{\begin{array}}
\newcommand\eeaa{\end{array}}
\newcommand{\mc}{\mathcal}

\def\XXint#1#2#3{{\setbox0=\hbox{$#1{#2#3}{\int}$ }
\vcenter{\hbox{$#2#3$ }}\kern-.5\wd0}}

\def\XXint#1#2#3{{\setbox0=\hbox{$#1{#2#3}{\int}$}
\vcenter{\hbox{$#2#3$}}\kern-.5\wd0}}

\newcommand{\nn}{\nonumber}

\newcommand{\neqa}{\nonumber\end{eqnarray}}
\newcommand{\la}[1]{\label{#1}}

\renewcommand{\P}{{\bf P}}

\newcommand{\eq}[1]{(\ref{#1})}

\newcommand{\Tr}{{\rm Tr}}

\newcommand{\hs}{\frac{\sqrt{3}}{2}}
\renewcommand{\d}{\partial}

\newcommand{\<}{{\langle}}
\renewcommand{\>}{{\rangle}}

\newcommand{\cD}{{\cal D}}

\newcommand{\re}{\relax{\rm I\kern-.18em R}}

\newcommand{\sh}{{\rm sh}}
\newcommand{\ch}{{\rm ch}}

\renewcommand{\sp}{p\hspace{-.40em}/}

\def\su2{{SU(2)}}

\def\eps{{\epsilon}}

\def\[{\left[}
\def\]{\right]}

\def\s{\sigma}

\def\({\left(}
\def\){\right)}
\def\[{\left[}
\def\]{\right]}

\def\<{\langle}
\def\>{\rangle}

\def\cO{{\cal O}}

\def\s*{\ *_{\!\!\!\!\!\!\!\!\!\,_{\,_\text{\scriptsize{sym}}}}}
\def\hs*{\ \hat{*}_{\!\!\!\!\!\!\!\!\!\,_{\,_\text{\scriptsize{sym}}}}}
\def\d{\partial}

\def\i2{\frac{i}{2}}

\def\bP{{\bf P}}

\def\spi{\relax{\rm \pi\kern-0.5em /}}
\def\sA{\relax{\rm A\kern-0.5em /}}
\def\sp{\relax{\rm p\kern-0.5em /}}
\def\sd{\relax{\rm \d\kern-0.5em /}}
\def\sk{\relax{\rm k\kern-0.5em /}}
\def\sn{\relax{\rm n\kern-0.5em /}}
\def\sl{\relax{\rm l\kern-0.5em /}}
\def\sP{\relax{\rm P\kern-0.7em /}}
\def\sBethe{\relax{\rm \Bethe\kern-0.5em /}}
\def\cN{{\cal N}}

	\newcommand{\mZ}{{\mathbb Z}}

	\newcommand{\rl}{\sqrt{\lambda}}
	\renewcommand{\Im}{{\rm Im}}
	
	\newcommand{\ofrac}[1]{\frac{1}{#1}}

\def\d{\partial}

    \newcommand{\AD}{AdS_5\times {\rm S}^5}

\title{
Quantum Spectral Curve at Work: From Small Spin to Strong Coupling
in ${\cal N}=4$ SYM
}

\author{Nikolay Gromov$^{1,2}$,
  Fedor Levkovich-Maslyuk$^{1}$,
  Grigory Sizov$^{1}$, Saulius Valatka$^{1}$ \\
  $^1$King's College London, Department of Mathematics, \\ The Strand, London WC2R 2LS,
  United Kingdom\\ \\
  $^2$ St.Petersburg INP, Gatchina, 188 300, St.Petersburg, Russia
  \qquad\\ \\
  \textit{E-mail:}
  \email{nikgromov$\bullet$gmail.com}, \email{fedor.levkovich$\bullet$gmail.com}, \email{grigory.sizov$\bullet$kcl.ac.uk },
\email{saulius.valatka$\bullet$kcl.ac.uk }\\ \\}

\abstract{
We apply the recently proposed quantum spectral curve technique to the study of twist operators in planar ${\cal N}=4$ SYM theory. We focus on the small spin expansion of anomalous dimensions in the sl(2) sector
and compute its first two orders exactly for any value of the `t Hooft coupling.
At leading order in the spin $S$ we reproduced Basso's slope function. The next term of order $S^2$ structurally
 resembles the Beisert-Eden-Staudacher dressing phase and takes into account wrapping contributions.
This expansion contains rich information about the
spectrum of local operators at strong coupling.
In particular, we found a new coefficient in the strong coupling expansion of the
Konishi operator dimension and confirmed several previously known terms.
We also obtained several new orders of the strong coupling expansion of the BFKL pomeron intercept.
As a by-product we formulated a prescription for the correct analytical continuation in $S$
which opens a way for deriving the BFKL regime of twist two anomalous dimensions from AdS/CFT integrability.

}

\preprint{}

\begin{document}

\section{Introduction}

Exploration of the holographic duality between planar 4D $\cN=4$ supersymmetric Yang-Mills theory (SYM) and string theory on $\AD$ has led to numerous remarkable results due to integrability discovered on both sides of the duality
\cite{Beisert:2010jr}. Integrability has been particularly successful in application to the problem of computing the
planar spectrum of single trace operator anomalous dimensions/string state energies. In the asymptotically large volume limit the spectrum was found to be captured by a system of nested asymptotic Bethe ansatz (ABA) equations \cite{Beisert:2005fw}. Finite-size corrections \cite{wrapping} were later accounted for via the Thermodynamic Bethe Ansatz (TBA)/Y-system technique
 \cite{Gromov:2009tv,Bombardelli:2009ns,Gromov:2009bc,Arutyunov:2009ur, Cavaglia:2010nm, Balog:2012zt}. This approach led to the formulation of an infinite set of integral equations, which are expected to describe the exact spectrum of the theory at any value of the 't Hooft coupling $\lambda$. The main problem of this
  approach is that the explicit form of the equations requires case-by-case study and is not known in general except for a
 few explicit examples such as Konishi \cite{Gromov:2009zb,Arutyunov:2012tx}. They, however, allowed for a detailed
 numerical study of these simplest operators \cite{Gromov:2009zb,Frolov:2010wt,Gromov:2011de,Frolov:2012zv} and led to a prediction for string theory which was confirmed in \cite{Roiban:2011fe,Vallilo:2011fj,Frolov:2013lva}.

Very recently a new set of equations called the quantum spectral curve or  the $\bP\mu$-system was proposed \cite{PmuPRL,PmuLong} which generalizes the original TBA equations to all sectors of the
theory and reveals a strikingly simple and concise underlying structure of the spectral problem. It allows one to describe all states of the theory on equal footing\footnote{This is in contrast to the analytic Y-system approach, which requires additional information about the
location of poles and/or zeros.
For simple states, like Konishi, these poles $u^*$ are prescribed to satisfy the
``exact Bethe ansatz equation" $Y_{1,0}^{\rm physical}(u^*)=-1$.
Already for more complicated states
in the $sl_2$ sector there are additional dynamical singularities in the Y-functions
which diverge from those appearing in the asymptotic solution when wrapping effects are taken into account.
$\bP\mu$-system puts under control all such singularities of $Y$-functions including those in even more complicated states.
In particular the BES equations with all types of the Bethe roots are a consequence of the $\bP\mu$-system (see \cite{PmuLong} for details).
The only input information it requires are the integer global R-charges and Lorentz spins entering
through the asymptotics of $\bP$-functions.
}.
The proposal has the form of a nonlinear Riemann-Hilbert problem for a set of a few functions.

Due to its remarkably transparent structure, the $\bP\mu$-system should be suitable to attack a variety of open problems
including such a longstanding problem of
AdS/CFT integrability as the description of the BFKL scaling regime.
Despite its novelty the $\bP\mu$-system was already used in various different situations.
One application which provided nontrivial tests of the proposal is the exact computation of
the Bremsstrahlung function \cite{PmuPRL,Gromov:2013qga}.
The new formulation also allowed to find the 9-loop Konishi anomalous dimension at weak coupling \cite{Volinnew}.
Below in the text we give a short overview of the construction
but we advice the reader to refer to \cite{PmuLong} where the quantum spectral curve is described in complete detail.

In this paper we will apply the $\bP\mu$-system to the calculation of twist operator anomalous dimensions in the $sl(2)$ sector of $\cN=4$ SYM. These operators have the form
\beq
	\cO=\Tr\(Z^{J-1}\;\cD^S Z\)+\dots
\eeq
where $Z$ denotes one of the scalars of the theory\footnote{Written in terms of two real scalars as $Z=\Phi_1+i\Phi_2$.}, $\cD$ is a lightcone covariant derivative and the dots stand for permutations. The number of derivatives $S$ is called the spin of the operator, while $J$ is called the twist. We will consider a two-cut configuration with a symmetric distribution of Bethe roots, thus for physical states $S$ is even. We will study the small spin limit, in which the scaling dimension of these operators can be written as
\beq
	\label{eq:anomalous_dimension_definition}
	\Delta = J+S+\gamma(g),\ \ \ \ g=\rl/(4\pi)
\eeq
with the anomalous dimension $\gamma(g)$ given as an expansion
\beq
	\label{eq:slope_definition}
	\gamma(g)=\gamma^{(1)}(g)S+\gamma^{(2)}(g)S^2+\mathcal{O}(S^3).
\eeq
The first term, $\gamma^{(1)}(g)$, is called the slope function. Remarkably, it can be found exactly at any value of the coupling \cite{Basso:2011rs}
\beq
\label{slopeIn}
	\gamma^{(1)}(g)=\frac{4\pi gI_{J+1}(4\pi g)}{JI_J(4\pi g)}\;.
\eeq
This expression was later derived from the ABA equations in two different ways \cite{Basso:2012ex,Gromov:2012eg}
and further studied and extended in \cite{Beccaria:2012kp,Beccaria:2012xm,Beccaria:2012mx,Tirziu:2008fk,Kruczenski:2012aw}.
This quantity is protected from finite-size wrapping corrections and thus the ABA prediction is exact. It is also not sensitive to the dressing phase of the ABA, which contributes only starting from order $S^2$.

Our key observation is that in the small $S$ regime the $\bP\mu$-system can be solved iteratively order by order in the spin. In this paper we first solve it at leading order and reproduce the slope function \eq{slopeIn}. Then we compute the coefficient of the $S^2$ term in the expansion, i.e. the function $\gamma^{(2)}(g)$ which we call the
\textit{curvature function}. For twist $J=2,3,4$ we obtain closed exact expressions for it in the form of a double integral. Unlike the slope function, $\gamma^{(2)}(g)$ is affected by the dressing phase in the ABA and by wrapping corrections, all of which
are incorporated in the exact $\bP\mu$-system.

Furthermore, we use the strong coupling expansion of our result to find the value of a new coefficient in the Konishi operator (i.e. $\Tr\(\cD^2Z^2\)$) anomalous dimension at strong coupling.
Our result for the Konishi dimension reads
\beq
	\Delta_{konishi}=2\,\lambda^{1/4}+\frac{2}{\lambda^{1/4}}+\frac{-3\,\zeta_3+\ofrac{2}}{\lambda^{3/4}}+\frac{\frac{15 \, \zeta_5}{2} + 6 \, \zeta_3+\frac{1}{2}}{\lambda^{5/4}}+\dots\ .
\eeq
We have also obtained two new terms in the strong coupling expansion of the BFKL pomeron intercept,
\beqa
j_0 = 2 + \left.S(\Delta)\right|_{\Delta=0} &=& 2 -\frac{2}{\lambda^{1/2}}-\frac{1}{\lambda }+ \frac{1}{4\,\lambda^{3/2}}+\left(6\,\zeta_3+2\right) \frac{1}{\lambda^2} \\
	&+& \nn \left(18 \, \zeta_3 + \frac{361}{64} \right) \frac{1}{\lambda^{5/2}} + \left(39 \, \zeta_3 + \frac{511}{32}\right) \frac{1}{\lambda^3}  + \mathcal{O}\left(\frac{1}{\lambda^{7/2}}\right),
\eeqa
where the new terms are in the second line.
In addition we have checked our results against available
results in literature at weak and strong coupling, and found full agreement.

The paper is organized as follows. First in section \ref{sec:pmu} we review the quantum spectral curve construction in a general setting. In section \ref{sec:exact_slope} we demonstrate its applicability by \text{rederiving} the exact slope function of $\mathcal{N}=4$ found
 in \cite{Basso:2011rs}. In section \ref{sec:exact_slope_to_slope} we push the calculation further and find the exact expression for the next coefficient in the small spin expansion,
i.e.
the curvature function. In sections \ref{sec:weak} and \ref{sec:SlopeSlopeStrongCoupling} we discuss the weak and strong coupling expansions of our result.
We then use our results to calculate the previously unknown three loop strong coupling coefficient of the Konishi anomalous dimension in subsection \ref{sec:Konishidimension} and two new coefficients for the BFKL intercept at strong coupling in subsection \ref{sec:bfkl}. We finish with conclusions and appendices, which contain detailed calculations left out of the main text for brevity.

\section{$\bP\mu$-system -- an overview}
\label{sec:pmu}

In this section we review the formulation of the $\bP\mu$-system, and also discuss its symmetries which will be useful later.
Below, we will restrict the discussion to states in the $sl(2)$ sector as presented in \cite{PmuPRL}.
Remarkably, the general case is not much more complicated and will appear soon in \cite{PmuLong}.

\subsection{Definitions and notation}

The $\bP\mu$-system is a nonlinear system of functional equations for a four-vector $\bP_a(u)$ and a $4\times4$ antisymmetric matrix $\mu_{ab}(u)$ depending on the spectral parameter $u$.
For full details about the origin of the construction we refer the reader to \cite{PmuLong}. As functions of $u$, both $\bP_a$ and $\mu_{ab}$ have prescribed analyticity properties which play a key role.
First, $\bP_a$ must have only a single branch cut in $u$ going between $-2g$ and $2g$, being analytic in the rest of the complex plane. We call this cut the \textit{short} cut, while the cut on the real
line connecting the same two points through infinity is called the \textit{long} cut. The functions $\mu_{ab}$ have an infinite set of short branch cuts going between $-2g+in$ and $2g+in$ for all $n\in\mZ$ (see Fig. \ref{fig:cuts}). Most importantly, the analytic continuation of $\bP_a$ and $\mu_{ab}$ through these cuts is again expressed in terms of these functions, according to the following equations:
\beq
\tilde \bP_a=-\mu_{ab}\chi^{bc}\bP_c,\; \ \ \ \text{with}\;\ \ \chi^{ab}=\left(
            \begin{array}{cccc}
              0 & 0 & 0 & -1 \\
              0 & 0 & 1 & 0 \\
              0 & -1 & 0 & 0 \\
              1 & 0 & 0 & 0 \\
            \end{array}
          \right),
\label{eq:Pmu}
\eeq
and
\beq
\tilde \mu_{ab}-\mu_{ab}=\bP_a \tilde\bP_b-  \bP_b \tilde\bP_a\;.
\label{eq:mudisc}
\eeq
Here we denote by $\tilde\bP_a$ and $\tilde\mu_{ab}$ the analytic continuation of $\bP_a$ and $\mu_{ab}$ through the cut on the real axis. In addition, we have a pseudo-periodicity condition
\beq
\label{muper}
	\tilde\mu_{ab}(u)=\mu_{ab}(u+i)
\eeq
which, actually, means that $\mu_{ab}(u)$ would be an $i$-periodic function if defined with long cuts instead of the short cuts.

\FIGURE[ht]
{
\label{fig:cuts}

    \begin{tabular}{cc}
    \includegraphics[scale=0.3]{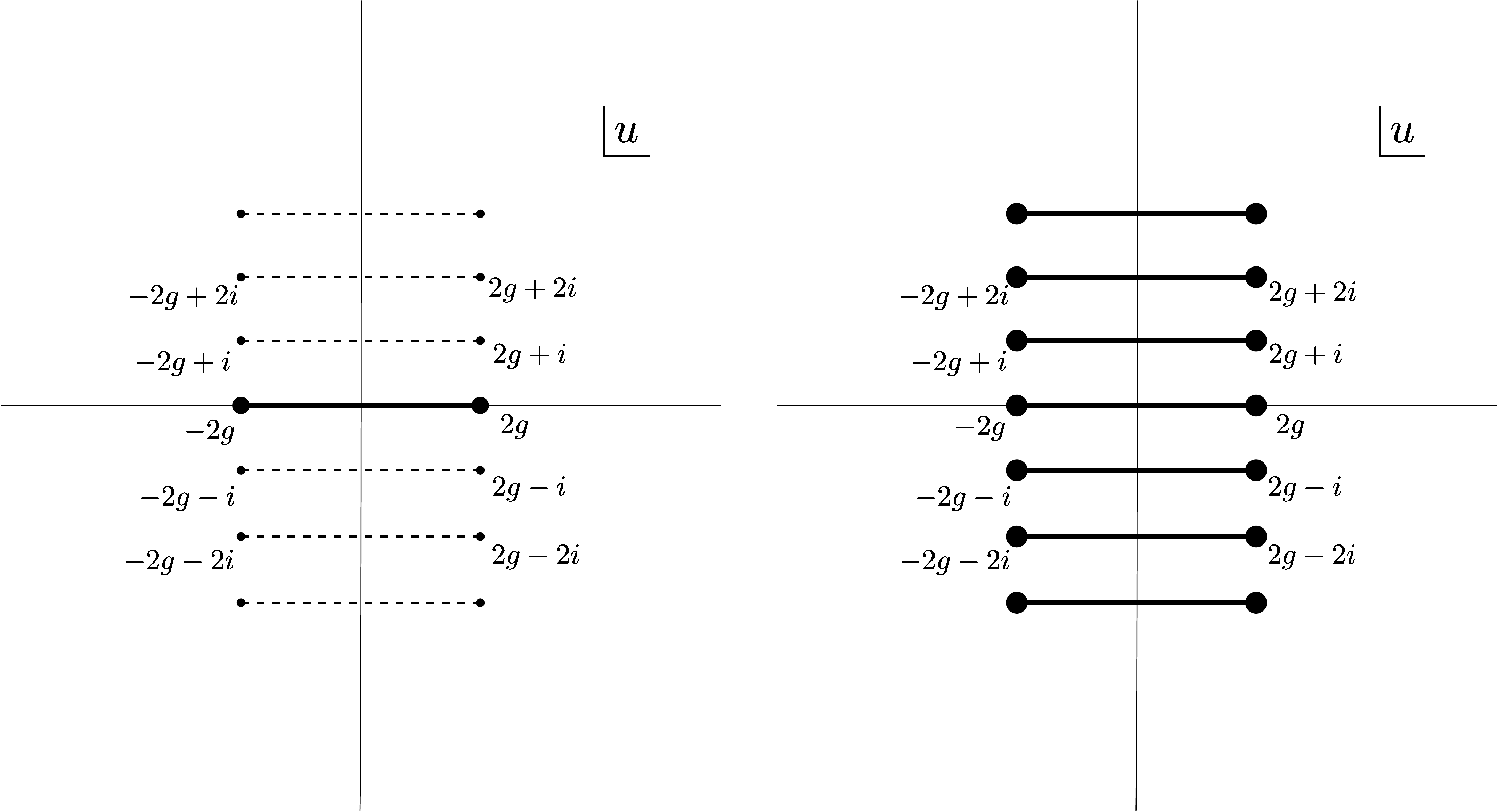}\\
    \end{tabular}
\caption{\textbf{Cuts in the $u$ plane.} We show the location of branch cuts in $u$ for the functions $\bP_a(u)$ (left) and $\mu_{ab}(u)$ (right). The infinitely many cuts of $\tilde\bP_a$ are shown on the left picture by dotted lines.}
}
The functions $\mu_{ab}$ are also constrained by the relations
\beqa
\label{constraint}
\mu_{12}\mu_{34}-\mu_{13}\mu_{24}+\mu_{14}^2&=&1\;,\\
\label{Pmulast}
\mu_{14}=\mu_{23}\;,
\eeqa
the first of which states that the Pfaffian
 of the matrix $\mu_{ab}$ is equal to $1$. Let us also write the equations \eq{eq:Pmu} explicitly:
\beqa
\label{eq:pmuexpanded1}
&&\tilde \bP_1= -\bP_3 \mu_{12}+\bP_2 \mu_{13}-\bP_1 \mu_{14} \\
&&\tilde \bP_2= -\bP_4 \mu_{12}\hspace{16mm}+\bP_2 \mu_{14}-\bP_1 \mu_{24} \\
&&\tilde \bP_3= \hspace{16mm}-\bP_4 \mu_{13}+\bP_3 \mu_{14}\hspace{16mm}-\bP_1 \mu_{34} \\
&&\tilde \bP_4= \hspace{16mm}\hspace{15.5mm}-\bP_4 \mu_{14}+\bP_3 \mu_{24}-\bP_2 \mu_{34}\;.
\label{eq:pmuexpanded}
\eeqa
The above equations ensure that the branch points of $\P_a$ and $\mu_{ab}$ are of the square root type, i.e.
$\tilde{\tilde{\bP}}_a=\bP_a$ and $\tilde{\tilde{\mu}}_{ab}=\mu_{ab}$.

Finally, we require that $\bP_a$ and $\mu_{ab}$ do not have any singularities except these branch points\footnote{For odd values of $J$ the functions $\bP_a$ may have an additional branch point at infinity. However, it should cancel in any product of two $\bP_a$'s, and therefore it will not appear in
any physically relevant quantity (see \cite{PmuPRL}, \cite{PmuLong}). We will discuss some explicit examples in the text.}.

\subsection{Asymptotics and energy}
The quantum numbers and the energy of the state are encoded in the asymptotics of the functions $\bP_a$ and $\mu_{ab}$ at large real $u$. The generic case is described in \cite{PmuLong}, while here we are interested in the states in the $sl(2)$ sector, for which the relations read \cite{PmuPRL}
\beq
\bP_a\sim(A_1u^{-J/2},A_2u^{-J/2-1},A_3u^{J/2},A_4u^{J/2-1})
\label{eq:asymptotics}
\eeq
\beq
	\(\mu_{12},\ \mu_{13},\ \mu_{14},\ \mu_{24},\ \mu_{34}\)\sim
	\(u^{\Delta-J},\ u^{\Delta+1},\ u^{\Delta},\ u^{\Delta-1},\ u^{\Delta+J}\)
\label{eq:muasymptotics}
\eeq
where $J$ is the twist of the gauge theory operator, and $\Delta$ is its conformal dimension. With these asymptotics, the equations \eq{eq:Pmu}-\eq{Pmulast} form a closed system which fixes $\bP_a$ and $\mu_{ab}$.

Lastly, the spin $S$ of the operator is related \cite{PmuPRL} to the leading coefficients $A_a$ of the $\bP_a$ functions (see \eq{eq:asymptotics}):
\beqa
&&A_1 A_4=\frac{\((J+S-2)^2-\Delta^2\)\((J-S)^2-\Delta^2\)}{16 i J(J-1)} \label{AA1} \\
&&A_2 A_3=\frac{\((J-S+2)^2-\Delta^2\)\((J+S)^2-\Delta^2\)}{16 i J(J+1)} \label{AA2}\;.
\eeqa

\subsection{Symmetries}
\label{sec:Symmetries}
The $\bP\mu$-system enjoys a symmetry preserving all of its essential features. It has the form of a linear transformation of $\bP_a$ and $\mu_{ab}$ which leaves the system \eqref{eq:Pmu}-\eqref{Pmulast} and the asymptotics \eq{eq:asymptotics}, \eq{eq:muasymptotics} invariant. Indeed, consider a general linear transformation $\bP_a'={R_a}^b \bP_b$ with a non-degenerate constant matrix $R$. In order to preserve the system \eqref{eq:Pmu}, $\mu$ should
at the same time be transformed as
\beq
\mu'=-R \mu \chi R^{-1}\chi.
\label{gammaP}
\eeq
Such a transformation also preserves the form of \eqref{eq:mudisc} if
\beq
R^T\chi R\chi=-1\;,
\label{eq:sxsx}
\eeq
which also automatically ensures antisymmetry of $\mu_{ab}$ and (\ref{constraint}), (\ref{Pmulast}).
In general, this transformation will spoil the asymptotics of $\bP_a$.
These asymptotics are ordered as $|\bP_2|<|\bP_1|<|\bP_4|<|\bP_3|$,
which implies that the   matrix $R$ must have the following structure\footnote{This matrix would of course be lower triangular if we ordered $\bP_a$ by their asymptotics.}
 \beq
R=\left(
\begin{array}{cccc}
 * & * & 0 & 0 \\
 0 & * & 0 & 0 \\
 * & * & * & * \\
 * & * & 0 & * \\
\end{array}
\right).
\eeq
The general form of $R$ which satisfies \eqref{eq:sxsx} and does not spoil the asymptotics generates a 6-parametric transformation, which we will call a $\gamma$-transformation. The simplest \text{$\gamma$-transformation} is the following rescaling:
\beq
\bP_1 \to \alpha \bP_1\;\;,\;\;
\bP_2 \to \beta \bP_2\;\;,\;\;
\bP_3 \to 1/\beta \bP_3\;\;,\;\;
\bP_4 \to 1/\alpha \bP_4\;\;,\;\;
\label{eq:alphabeta}
\eeq
\beq
\mu_{12} \to \alpha\beta\mu_{12}\;\;,\;\;
\mu_{13} \to \frac{\alpha}{\beta}\mu_{13}\;\;,\;\;
\mu_{14} \to \mu_{14}\;\;,\;\;
\mu_{24} \to \frac{\beta}{\alpha}\mu_{24}\;\;,\;\;
\mu_{34} \to \frac{1}{\alpha\beta}\mu_{34}\;\;,\;\;
\eeq
with $\alpha,\beta$ being constants.

 In all the solutions we consider in this paper all functions $\bP_a$ turn out to be functions of definite parity, so it makes sense to consider $\gamma$-transformations which preserve parity. $\bP_1$ and $\bP_2$  always have opposite parity (as one can see from from \eqref{eq:asymptotics}) and thus should not mix under such transformations; the same is true about $\bP_3$ and $\bP_4$. Thus, depending on parity of $J$ the parity-preserving $\gamma$-transformations are either

\beqa
\label{gammatransform2}
&\bP_3\rightarrow\bP_3+\gamma_3\bP_2,\ \bP_4\rightarrow\bP_4+\gamma_2\bP_1,\\
\nn&\mu_{13}\rightarrow\mu_{13}+\gamma_3\mu_{12},\ \mu_{24}\rightarrow\mu_{24}-\gamma_2\mu_{12},\ \mu_{34}\rightarrow\mu_{34}+\gamma_3\mu_{24}-\gamma_2\mu_{13}-\gamma_2\gamma_3\mu_{12}
\eeqa
for odd $J$ or
\beqa
\label{gammatransform1}
&\bP_3\rightarrow\bP_3+\gamma_1\bP_1,\ \bP_4\rightarrow\bP_4-\gamma_1\bP_2,\\
\nn&\mu_{14}\rightarrow\mu_{14}-\gamma_1\mu_{12},\ \mu_{34}\rightarrow\mu_{34}+2\gamma_1\mu_{14}-\gamma^2_1\mu_{12}\;,
\eeqa
for even $J$.

\section{Exact slope function from the $\bP\mu$-system}
\label{sec:exact_slope}

In this section we will find the solution of the $\bP\mu$-system \eqref{eq:Pmu}-\eq{Pmulast} corresponding to the $sl(2)$ sector operators at leading order in small $S$. Based on this solution we will compute the slope function $\gamma^{(1)}(g)$ for any value of the coupling.

\subsection{Solving the $\bP\mu$-system in LO}
\label{sec:evenLsol}
The solution of the $\bP\mu$-system is a little simpler for even $J$, because for odd $J$ extra branch points at infinity will appear in $\bP_a$ due to the asymptotics \eq{eq:asymptotics}. Let us first consider the even $J$ case.

The description of the $\bP\mu$-system in the previous section was done for physical operators. Our goal is to take some peculiar limit
when the (integer) number of covariant derivatives
$S$ goes to zero.  As we will see this requires some extension of the asymptotic requirement for $\mu$ functions.
 In this section we will be guided by principles of naturalness and simplicity to deduce these modifications which
 we will summarize in section~\ref{sec:ancont}. There we also give a concrete prescription for analytical continuation in $S$, which we then use to derive the curvature function.

We will start by finding $\mu_{ab}$. Recalling that $\Delta=J+{\cal O}(S)$, from \eq{AA1}, \eq{AA2} we see that $A_1A_4$ and $A_2A_3$ are of order $S$ for small $S$, so we can take the functions $\bP_a$ to be of order $\sqrt{S}$. This is a key simplification,
because now \eq{eq:mudisc} indicates that the discontinuities of $\mu_{ab}$ on the cut are small when $S$ goes to zero. Thus at leading order in $S$ all $\mu_{ab}$ are just periodic entire functions without cuts.
For power-like asymptotics of $\mu_{ab}$ like in \eq{eq:muasymptotics} the only possibility is that they are all constants.
However, we found that in this case there is only a trivial solution, i.e. $\bP_a$ can only be zero.
The reason for this is that for physical states $S$ must be integer and thus cannot be arbitrarily small, nevertheless, it is a sensible
 question how to define an analytical continuation from integer values of $S$.\footnote{Restricting the large positive $S$ behavior
 one can achieve uniqueness of the continuation.}

Thus we have to relax the requirement of power-like behavior at infinity. The first possibility is
to allow for $e^{2\pi u}$ asymptotics at $u\to +\infty$.
We should, however, remember about the constraints \eq{constraint} and \eq{Pmulast} which restrict our choice and the fact that we can also use $\gamma$-symmetry.
Let us show that by allowing $\mu_{24}$ to have exponential behavior and setting it to
$\mu_{24}=C\sinh(2\pi u)$, with other $\mu_{ab}$ being constant, we arrive to the correct result. This choice is dictated by our assumptions concerning the analytic continuation of $\mu_{ab}$ to non-integer values of $S$, and this point is discussed in detail in section~\ref{sec:ancont}. We will also see in that section that by using the $\gamma$-transformation (described in section \ref{sec:Symmetries}) and the constraint \eq{constraint} we can set the constant $C$ to $1$ and also $\mu_{12}=1,\;\mu_{13}=0,\;\mu_{14}=-1,\;\mu_{34}=0$ (see \eq{muresan}).
%

Having fixed all $\mu$'s at leading order we get the following system of equations\footnote{In this section we only consider the leading order of $\bP$'s at small $S$, so the equations involving them are understood to hold at leading order in $S$. In section 4 we will study the next-to-leading order and elaborate the notation for contributions of different orders.} for $\bP_a$:
\beqa
&&\tilde \bP_1= -\bP_3 +\bP_1, \label{eq:P1L2} \\
&&\tilde \bP_2= -\bP_4 -\bP_2 -\bP_1 \sinh(2\pi u), \label{eq:P2L2}\\
&&\tilde \bP_3= \hspace{10mm}-\bP_3,\hspace{16mm} \label{eq:P3L2} \\
&&\tilde \bP_4= \hspace{10mm}+\bP_4+\bP_3 \sinh(2\pi u).\label{eq:P4L2}
\eeqa
Recalling that the functions $\bP_a$ only have a single short cut, we see from these equations that $\tilde{\bP}_a$ also have only this cut! This means that we can take all $\bP_a$ to be infinite Laurent series in the Zhukovsky variable $x(u)$, which rationalizes the Riemann surface with two sheets and one cut. It is defined as
\beq
	x+\ofrac{x}=\frac{u}{g}
\eeq
where we pick the solution with a short cut, i.e.
\beq
x(u)=\frac{1}{2}\left(\frac{u}{g}+\sqrt{\frac{u}{g}-2}\;\sqrt{\frac{u}{g}+2}\;\right)\;\;.\;\;
\eeq
Solving the equation \eqref{eq:P3L2} with the asymptotics \eqref{eq:asymptotics} we find
\beq
\label{P3ck}
	\bP_3=\epsilon\(x^{-J/2}-x^{+J/2}\)+\sum_{k=1}^{J/2-1}c_k\(x^{-k}-x^k\)
\eeq
where $\epsilon$ and $c_k$ are constants. Now it is useful to rewrite the equation for $\bP_1$ (i.e. \eq{eq:P1L2}) in the form $\tilde\bP_1-\bP_1=-\bP_3$, and we see that due to asymptotics of $\bP_1$ both sides of this equation must have a gap in the powers of $x$ from $x^{-J/2+1}$ to $x^{J/2-1}$. This means that all coefficients $c_k$ in \eq{P3ck} must vanish and we find
\beq
	\bP_1=\epsilon x^{-J/2} \ ,
\eeq
so we are left with one unfixed constant $\epsilon$ (we expect it to be proportional to $\sqrt{S}$).


Thus the equations \eqref{eq:P2L2} and \eqref{eq:P4L2} become
\beqa
\label{eq:P2eq}
\tilde \bP_2+\bP_2&=& -\bP_4 -\epsilon x^{-J/2}\sinh(2\pi u)\;, \\
\label{eq:P4eq}
\tilde \bP_4-\bP_4&=& \epsilon(x^{-J/2}-x^{+J/2}) \sinh(2\pi u)\;.
\eeqa
We will first solve the second equation.
It is useful to introduce operations $[f(x)]_+$ and $[f(x)]_-$, which take parts of Laurent series with positive and negative powers of $x$ respectively.  Taking into account that
\beq
	\sinh(2\pi u)=\sum\limits_{n=-\infty}^{\infty}I_{2n+1}  x^{2n+1},
\eeq
where $I_k\equiv I_{k}(4 \pi g)$ is the modified Bessel function of the first kind, we can write $\sinh(2\pi u)$ as

 \beq
 \sinh(2\pi u)= \sinh_++\sinh_-,
 \eeq
 where explicitly
 \beqa
&& \sinh_+=[\sinh(2\pi u)]_+=\sum\limits_{n=1}^\infty I_{2n-1}x^{2n-1} \\
\label{defshm}
&& \sinh_-=[\sinh(2\pi u)]_-=\sum\limits_{n=1}^\infty I_{2n-1}x^{-2n+1}\;.
 \eeqa
In this notation the general solution of Eq. \eq{eq:P4eq} with asymptotics at infinity $\bP_4\sim u^{J/2-1}$ can be written as
\beq
\bP_4=\epsilon(x^{J/2}-x^{-J/2})\sinh_-+Q_{J/2-1}(u),
\eeq
where $Q_{J/2-1}$ is a polynomial of degree $J/2-1$ in $u$.
The polynomial $Q_{J/2-1}$ can be fixed from the equation \eqref{eq:P2eq} for $\bP_2$. Indeed, from the asymptotics of $\bP_2$ we see that the lhs of \eqref{eq:P2eq} does not have powers of $x$ from $-J/2+1$ to $J/2-1$. This fixes
\beq
Q_{J/2-1}(x)=-\epsilon\sum\limits_{k=1}^{J/2}I_{2k-1}\(x^{\frac{J}{2}-2k+1}+x^{-\frac{J}{2}+2k-1}\).
\eeq
Once $Q_{J/2-1}$ is found, we set $\bP_2$ to be the part of the right hand side of \eqref{eq:P2eq} with powers of $x$ less than $-J/2$, which gives
\beq
\bP_2=-\epsilon x^{+J/2} \sum_{n=\frac{J}{2}+1}^\infty I_{2n-1}x^{1-2n}.
\eeq
Thus (for even $J$) we have uniquely fixed all $\bP_a$ with the only unknown parameter being $\epsilon$. We summarize the solution below:
\beqa
\label{eq:musolLOevenL}
&&\mu_{12}=1,\ \mu_{13}=0,\ \mu_{14}=-1,\ \mu_{24}=\sinh(2\pi u),\ \mu_{34}=0,\\
\label{eq:P1solLOevenL}
&&\bP_1=\epsilon x^{-J/2}\\
\label{eq:P2solLOevenL}
&&\bP_2=-\epsilon x^{+J/2} \sum_{n={J/2}+1}^\infty I_{2n-1}x^{1-2n}\\
\label{eq:P3solLOevenL}
&&\bP_3=\epsilon \(x^{-J/2}-x^{+J/2}\)\\
\label{eq:P4solLOevenL}
&&
	\bP_4=\epsilon \(x^{J/2}-x^{-J/2}\)\sinh_- -\epsilon \sum\limits_{n=1}^{J/2}I_{2n-1}\(x^{\frac{J}{2}-2n+1}+x^{-\frac{J}{2}+2n-1}\)\;.
\label{solutionevenL}
\eeqa
In the next section we fix the remaining parameter $\epsilon$ of the solution in terms of $S$ and find the energy, but now
let us briefly discuss the solution for odd $J$. As we mentioned above the main difference is that the functions $\bP_a$ now have a branch point at $u=\infty$, which is dictated by the asymptotics \eq{eq:asymptotics}. In addition, the parity of $\mu_{ab}$ is different according to the asymptotics of these functions \eq{eq:muasymptotics}. The solution is still very similar to the even $J$ case, and we discuss it in detail in Appendix \ref{sec:oddL}. Let us present the result here:
\beq
	\mu_{12}=1,\ \mu_{13}=0,\ \mu_{14}=0,\  \mu_{24}=\cosh(2\pi u),\ \mu_{34}=1
\eeq
\beqa
\label{P1oddL}
&&   \bP_1=\epsilon  x^{-J/2}, \\
&&   \bP_2=-\epsilon  x^{J/2}\sum\limits_{k=-\infty}^{-\frac{J+1}{2}}I_{2k}x^{2k},\\
&&   \bP_3=-\epsilon  x^{J/2}, \\
\label{P4oddL}
&&    \bP_4=\epsilon  x^{-J/2}\cosh_--\epsilon  x^{-J/2}\sum\limits_{k=1}^{\frac{J-1}{2}}I_{2k}x^{2k}-\epsilon  I_0 x^{-J/2}.
\eeqa
Note that now $\bP_a$ include half-integer powers of $x$.

\paragraph{Fixing the global charges of the solution.}
\label{sec:LOresultevenL}
Finally, to fix our solution completely we
have to find the value of $\epsilon$ and find the energy in terms of the spin using \eqref{AA1} and \eq{AA2}.
 For this we first extract the coefficients $A_a$ of the leading terms for all $\bP_a$ (see the asymptotics \eqref{eq:asymptotics}).
From \eqref{eq:P1solLOevenL}-\eqref{eq:P4solLOevenL} or \eq{P1oddL}-\eq{P4oddL}
we get
\beqa
\label{Aexp1}
&& A_1= g^{J/2} \epsilon , \\
&& A_2=-g^{J/2+1} \epsilon  I_{J+1}, \\
\label{eq:A3LOL3}
&& A_3=-g^{-J/2} \epsilon , \\
\label{Aexplast}
&& A_4=-g^{-J/2+1}\epsilon  I_{J-1}.
\eeqa
Expanding \eq{AA1}, \eq{AA2} at small $S$ with $\Delta=J+S+\gamma$, where $\gamma={\cal O}(S)$, we find at linear order
\beqa
&& \gamma=i(A_1 A_4-A_2 A_3) \\
&& S=i(A_1A_4+A_2A_3)\;.
\eeqa
Plugging in the coefficients \eq{Aexp1}-\eq{Aexplast} we find that
\beq
\label{epss}
	\epsilon=\sqrt{\frac{2\pi i S}{JI_J(\sqrt\lambda)}}
\eeq
and we obtain the anomalous dimension at leading order,
\beq
\gamma=\frac{\sqrt{\lambda}I_{J+1}(\sqrt{\lambda})}{JI_J(\sqrt{\lambda})}S+{\cal O}(S^2),
\label{eq:resultLO}
\eeq
which is precisely the slope function of Basso \cite{Basso:2011rs}.

While the above discussion concerned the ground state, i.e. the $sl(2)$ sector operator with the lowest anomalous dimension at given twist $J$, it can be
generalized for higher mode numbers. In the asymptotic Bethe ansatz for such operators we have two symmetric cuts formed by Bethe roots, with corresponding mode numbers being $\pm n$ (for the ground state $n=1$). To describe these operators within the $\bP\mu$-system
we found that we should take $\mu_{24}=C\sinh(2\pi n u)$ instead of $\mu_{24}=C\sinh(2\pi u)$ (and for odd $J$ we similarly use $\mu_{24}=C\cosh(2\pi n u)$ instead of $\mu_{24}=C\cosh(2\pi u)$). Then the solution is very similar to the one above, and we find
\beq
\gamma=\frac{n\sqrt{\lambda}I_{J+1}(n\sqrt{\lambda})}{JI_J(n\sqrt{\lambda})}S\;,
\label{slopen}
\eeq
which reproduced the result of \cite{Basso:2011rs} for  non-trivial mode number $n$. In Appendix \ref{sec:Sanyn} we also show how using the $\bP\mu$-system one can reproduce the slope function for a configuration of Bethe roots with arbitrary mode numbers and filling fractions.

In summary, we have shown how the $\bP\mu$-system correctly computes  the energy at linear order in $S$. In section \ref{sec:exact_slope_to_slope} we will compute the next, $S^2$ term in the anomalous dimension.

\subsection{Prescription for analytical continuation}
\label{sec:ancont}
To deduce the general prescription for the asymptotics of $\mu_{ab}$ for non-integer $S$ from our analysis,
we first study the possible asymptotics of $\mu_{ab}$
for given $\bP_a$ in more detail. For that we combine \eq{muper}
with \eq{eq:mudisc} and \eq{eq:Pmu} to write a finite difference equation on $\mu_{ab}$:
\beq\la{5bax}
\mu_{ab}(u+i)=\mu_{ab}(u)-\mu_{bc}(u)\chi^{cd}\bP_d\bP_a+\mu_{ac}(u)\chi^{cd}\bP_d\bP_b.
\eeq
As there are $5$ linear independent components of $\mu_{ab}$ this is a 5th order
finite-difference equation which has $5$ independent solutions which we denote $\mu_{ab,A},\;A=1,\dots,5$.
 Given the asymptotics of $\bP_a$ \eq{eq:asymptotics} and \eq{AA1}, \eq{AA2}
there are exactly $5$ different asymptotics a solution of \eq{5bax} could have as discussed in \cite{PmuPRL}.
We denote these $5$ independent solutions of \eq{5bax} as $\mu_{12,A}$ where $A=1,\dots,5$ and summarize their
leading asymptotics at large $u>0$ in the table below
\beq
\bea{c||l|l|l|l|l}
A=&1&2&3&4&5\\ \hline\hline
\mu_{12,A}\sim & u^{\Delta-J}& C_{1,2} u^{-S+1-J}& C_{1,3} u^{-J}& C_{1,4} u^{S-1-J}& C_{1,5} u^{-\Delta-J}\\
\mu_{13,A}\sim & C_{2,1} u^{\Delta+1}& C_{2,2} u^{-S+2}& C_{2,3} u^{+1}& { u^{S}}& C_{2,5} u^{-\Delta+1}\\
\mu_{14,A}\sim & C_{3,1} u^{\Delta}& C_{3,2} u^{-S+1}& 1 & C_{3,4} u^{S-1}& C_{3,5} u^{-\Delta}\\
\mu_{24,A}\sim & C_{4,1} u^{\Delta-1}& u^{-S}& C_{4,3} u^{-1}& C_{4,4} u^{S-2}& C_{4,5} u^{-\Delta-1}\\
\mu_{34,A}\sim & C_{5,1} u^{\Delta+J}& C_{5,2} u^{-S+1+J}& C_{5,3} u^{+J}& C_{5,4} u^{S-1+J}& { u^{-\Delta+J}}
\label{tablemu}
\eea
\eeq
where we fix the normalization of our solutions so that some coefficients are set to $1$\footnote{The coefficients $C_{a,A}$ are some rational
functions of $S,\Delta,J$ and $A_1,A_2$. In the small $S$ limit all $C_{a,A}\to 0$ in our normalization.}.
As it was pointed out in \cite{PmuPRL} the asymptotics for different $A's$ are obtained by replacing $\Delta$ in \eq{eq:muasymptotics} by $\pm \Delta,\pm (S-1) $ and $0$.
We label these solutions so that in the small $S$ regime these asymptotics are ordered $\Delta> 1-S>0>S-1>-\Delta$.

Of course any solution of \eq{5bax} multiplied by an $i$-periodic function\footnote{It could be a periodic function with short cuts. In general the set of these coefficients is denoted in
\cite{PmuLong} by $\omega_{ab}$ whereas $\mu_{ab,A}$ is denoted in \cite{PmuLong} as ${\cal Q}_{ab,cd}$.} will still remain a solution of \eq{5bax}.
The true $\mu_{ab}$ is thus a linear combination of the partial solutions $\mu_{ab,A}$ with some constant or periodic coefficients.
This particular combination should in addition satisfy the analyticity condition \eq{muper} which is not guaranteed by \eq{5bax}.

The prescription for analytical continuation in $S$ which we propose here is based on the large $u$ asymptotics of these periodic coefficients.
As we discussed in the previous section the assumption that all these coefficients are asymptotically constant
is too constraining already at the leading order in $S$, and we must assume that at least some of these coefficients grow exponentially  as $e^{2\pi u}$.
To get some extra insight into the asymptotic behavior of these coefficients it is very instructive to go to the weak coupling regime.

It is known that at one loop the equation \eq{5bax} reduces to a second order equation. When written as a finite difference equation for $\mu_{12}$ it coincides
exactly with the Baxter equation for the non-compact $sl(2)$ spin chain. For $J=2$ it reads
\beq\la{oneloopbaxter}
\(2u^2-S^2-S-\frac{1}{2}\)Q(u)=(u+\tfrac{i}{2})^2 Q(u+i)
+(u-\tfrac{i}{2})^2 Q(u-i)
\eeq
where $Q(u)=\mu_{12}(u+i/2)$.
This equation is already very well studied and all its solutions are known explicitly \cite{Derkachov:2002wz} -- in particular it is easy to see that one of the solutions
must have $u^S$ asymptotics at infinity, while the other behaves as $1/u^{S+1}$.
It is also known that at one loop and for any integer $S$ \eq{oneloopbaxter} has a polynomial solution which gives the energy as $\Delta=J+S+
\left.2ig^2\d_u\log\frac{Q(u-i/2)}{Q(u-i/2)}\right|_{u=0}=S+J+8g^2 H_S$.
At the same time, for non-integer $S$ there are of course no polynomial solutions,
and according to \cite{Janik:2013nqa} and \cite{GrKazakov}
the solution which produces the energy $S+J+8g^2 H_S$
cannot even have power-like asymptotics, instead the correct large $u$ behavior must be:
\beq
Q(u)\sim \(u^S+\dots\)+(A+B e^{2\pi u})\(\frac{1}{u^{S+1}}+\dots\)\;\;,\;\;u\to +\infty\;.
\eeq
Furthermore, there is a unique entire $Q$ function with the above asymptotics.
For $S>-1/2$ we can reformulate the prescription by saying that the correct solution
has power-like asymptotics, containing all possible solutions, plus a small solution reinforced with an exponent.

In this form we can try to translate this result to our case. We notice that for $g\to 0$ we have
 $\mu_{12,1}\sim u^{S}$ and $\mu_{12,2}\sim u^{-S-1}$, which tells us that at least the second solution
 must be allowed to have a non-constant periodic coefficient in the asymptotics. We also assume that the coefficient in front of $\mu_{ab,3}$ tends to a constant\footnote{It could be hard or even impossible to separate $\mu_{ab,3}$ from $\mu_{ab,2}$ in a well defined way.
 In these cases $\mu_{ab,2}$ is defined modulo $\mu_{ab,3}$ and other subleading solutions. Our prescription then means that the exponential part of the coefficient in front of $\mu_{ab,3}$ is proportional to that of in front of $\mu_{ab,2}$.
 }.
 This extra condition does not follow from the one loop analysis we deduced from our solution.
 We will show how this prescription produces the correct known result for the leading order in $S$.
 From our analysis it is hard to make a definite statement about the behavior of the periodic coefficients in front of
 $\mu_{12,4}$ and  $\mu_{12,5}$, but due to the expected $\Delta \to-\Delta$ symmetry, which interchanges $\mu_{12,5}$ and $\mu_{12,1}$,
 one may expect that the coefficient of $\mu_{12,5}$ should also go to a constant. To summarize we should have
\beq\la{prescr}
\mu_{ab}(u)=\sum_{A=1}^5 c_{A}\mu_{ab,A}(u)
+\sum_{A=2,4,5} p_{A}(u)\mu_{ab,A}(u)
\eeq
where $c_A$ are constants whereas $p_{A}(u)$ are some linear combinations of $e^{\pm 2\pi u}$.\footnote{It could be that some of the coefficients of $p_{A}$ should be zero due to the constraint \eq{constraint}.}

\paragraph{Prescription at small $S$.}
In the small $S$ limit, $\bP_a\to 0$ and the equation \eq{5bax} simply tells us that $\mu_{ab}(u+i)=\mu_{ab}(u)$ which implies that our $5$ independent solutions are
just constants at the leading order in $S$.
We begin by noticing that in this limit $\mu_{12}$ must be entirely coming from $\mu_{12,1}$ as all the other solutions could only produce negative powers and thus cannot
contribute at the leading order. So we start from $\mu_{ab}=C_{ab}+D_{ab}\sinh(2\pi u)+E_{ab}\cosh(2\pi u)$ for some constants $C_{ab},D_{ab},E_{ab}$ such that
$D_{12}=E_{12}=0$. Thus we have $5$ different $C$'s, $4$ different  $D$'s and $4$ different $E$'s.
We notice that this general form of $\mu_{ab}$ can be significantly simplified.
First, using the Pfaffian constraint \eqref{constraint} and the \text{$\gamma$-transformation} \eqref{gammaP} any generic $\mu_{ab}$ of this form
can be reduced to one belonging to the following two-parametric family inside the original 13-parametric space:
\beqa
&&\mu_{12}=1,\ \mu_{14}=a^2 \sinh{2\pi u}+\frac{a}{2}\cosh{2\pi u}\;,\\\
&&\mu_{24}=b\sinh{2\pi u}+\sinh{2\pi u}\;,\
\mu_{34}=\frac{a^2}{4}\frac{(1-2ab)^2}{b^2-1}+1\;,
\label{mufamilyab}
\eeqa
where $\mu_{13}$ is found from the Pfaffian constraint.
Second, recall that according to our prescription the 1st and 3rd solutions (columns in the table \eqref{tablemu}) cannot contain exponential terms.
Consider $\mu_{14}$ and $\mu_{24}$, we again see that the 4th and 5th solutions
could only contain negative powers of $u$ and thus only the 2nd solution can contribute to the parts of $\mu_{14}$ and $\mu_{24}$ that are non-decaying at infinity.
This means that these components can be represented in the following form
\beqa
\mu_{14}=(a_1\sinh{2\pi u}+ a_2\cosh{2\pi u})\mu_{14,2}(u)+\mathcal{O}\(e^{2\pi u}/u\)\;,\\
\mu_{24}=(a_1\sinh{2\pi u}+a_2\cosh{2\pi u})\mu_{24,2}(u)+\mathcal{O}\(e^{2\pi u}/u\)\;,
\eeqa
for $u\to+\infty$.
The  $\mathcal{O}\(e^{2\pi u}/u\)$ terms contain contributions from all of the solutions except for the 2nd. One can see that \eqref{mufamilyab} can be of this form only in two cases: if $a=0$ or if $a=\frac{1}{2b}$.
Both of these cases can be brought to the form
\beq
\label{muresan}
\mu_{12}=1,\ \mu_{13}=0,\ \mu_{14}=0,\ \mu_{24}=d_1\sinh{2\pi u}+d_2\cosh{2\pi u},\ \mu_{34}=1
\eeq
by a suitable $\gamma$-transformation \eqref{mufamilyab}. However, we found that there is an additional constraint which follows from compatibility of $\mu_{ab}$
with the decaying asymptotics of $\bP_2$. As
we show in appendix \ref{sec:Sanyn} for even $J$ one must set $d_2=0$. For odd $J$ we must set $d_1=0$ as a compatibility
requirement. This justifies the choice of $\mu_{ab}$ used in the previous section.
In the next section we will show how the same prescription can be applied at the next order in $S$ and leads to nontrivial results which
we subjected to intensive tests later in the text.

\section{Exact curvature function}
\label{sec:exact_slope_to_slope}

In this section we use the $\bP\mu$-system to compute the $S^2$ correction to the anomalous dimension, which we call the curvature function $\gamma^{(2)}(g)$. First we will discuss the case $J=2$ in detail and then describe the modifications of the solution for the cases $J=3$ and $J=4$, more details on which can be found in appendix \ref{sec:NLOapp}.

\subsection{Iterative procedure for the small $S$ expansion of the $\bP\mu$-system}
\label{sec:SolvingPmuL2}

For convenience let us repeat the leading order solution of the $\bP\mu$-system for $J=2$ (see \eqref{eq:musolLOevenL}-\eqref{eq:P4solLOevenL})
\beqa
{\bf P}^{(0)}_1=\epsilon\frac{1}{x}\;\;&,&\;\;{\bf P}^{(0)}_2=+\epsilon I_1-\epsilon x[\sinh(2\pi u)]_-\;\;,\\
{\bf P}^{(0)}_3=\epsilon\(\frac{1}{x}-x\)\;\;&,&\;\;
{\bf P}^{(0)}_4=
-2\epsilon I_1-
\epsilon \(\frac{1}{x}-x\)[\sinh(2\pi u)]_-.
\label{P10P40}
\eeqa
Here $\epsilon$ is a small parameter, proportional to $\sqrt{S}$ (see \eq{epss}), and by $\bP_a^{^{(0)}}$ we denote the $\bP_a$ functions at leading order in $\epsilon$.

The key observation is that the $\bP\mu$-system can be solved iteratively order by order in $\epsilon$. Let us write $\bP_a$ and $\mu_{ab}$ as an expansion in this small parameter:
\beq
	\bP_a=\bP_a^{(0)}+\bP_a^{(1)}+\bP_a^{(2)}+\dots
\eeq
\beq
	\mu_{ab}=\mu_{ab}^{(0)}+\mu_{ab}^{(1)}+\mu_{ab}^{(2)}+\dots \;.
\eeq
where $\bP_a^{(0)}=\cO(\eps),\;\bP_a^{(1)}=\cO(\eps^3),\;\bP_a^{(2)}=\cO(\eps^5),\;\dots$, and $\mu_{ab}^{(0)}=\cO(\eps^0),\; \mu_{ab}^{(1)}=\cO(\eps^2),\; \mu_{ab}^{(2)}=\cO(\eps^4),$ etc.
This structure of the expansion is dictated by the equations \eq{eq:Pmu}, \eq{eq:mudisc} of the $\bP\mu$-system (as we will soon see explicitly). Since the leading order $\bP_a$ are of order $\epsilon$, equation \eqref{eq:mudisc} implies that the discontinuity of $\mu_{ab}$ on the cut is of order $\epsilon^2$. Thus to find $\mu_{ab}$ in the next to leading order (NLO) we only need the functions $\bP_a$ at leading order. After this, we can find the NLO correction to $\P_a$ from equations \eqref{eq:mudisc}. This will be done below, and having thus the full solution of the $\bP\mu$-system at NLO we will find the energy at order $S^2$.

\subsection{Correcting $\mu_{ab}$\dots}
\label{sec:muNLOL2}
In this subsection we find the NLO corrections $\mu^{(1)}_{ab}$ to $\mu_{ab}$. As follows from \eqref{eq:mudisc} and \eq{muper},
they should satisfy the equation
\beq
 \mu_{ab}^{(1)}(u+i)-\mu_{ab}^{(1)}(u)=\bP_a^{(0)} \tilde\bP_b^{(0)}-  \bP_b^{(0)} \tilde\bP_a^{(0)},
\label{eq:mudiscNLO}
\eeq
in which the right hand is known explicitly. For that reason let us define an apparatus for solving equations of this type, i.e.
\beq
f(u+i)-f(u)=h(u).
\label{eqperiod}
\eeq
More precisely, we consider functions $f(u)$ and $h(u)$ with one cut in $u$ between $-2g$ and $2g$, and no poles. Such functions can be represented as infinite Laurent series in the Zhukovsky variable $x(u)$, and we additionally restrict ourselves to the case where for $h(u)$ this expansion does not have a constant term\footnote{The r.h.s. of \eq{eq:mudiscNLO} has the form $F(u)-\tilde F(u)$ and therefore indeed does not have a constant term in its expansion, as the constant in $F$ would cancel in the difference $F(u)-\tilde F(u)$.}.

One can see that the general solution of \eqref{eqperiod} has a form of a particular solution plus an arbitrary $i$-periodic function, which we also call a zero mode. 
 First we will describe the construction of the particular solution and later deal with zero modes. The linear operator which gives the particular solution of \eqref{eqperiod} described below will be denoted as $\Sigma$.

Notice that given the explicit form \eqref{P10P40} of $\bP^{(0)}_a$, the right hand side of \eqref{eq:mudiscNLO} can be represented in a form
\beq
\alpha(x)\sinh(2\pi u)+\beta(x),
\label{alphabetasinh}
\eeq
 where $\alpha(x),\beta(x)$ are power series in $x$  growing at infinity not faster than polynomially. Thus for such $\alpha$ and $\beta$ we define
\beq
\Sigma\cdot\[\alpha(x)\sinh(2\pi u)+\beta(x)\]\equiv \sinh(2\pi u) \Sigma\cdot \alpha(x)+\Sigma\cdot \beta(x).
\eeq
We also define $\Sigma\cdot x^{-n}=\Gamma'\cdot x^{-n}$ for $n>0$, where the integral operator $\Gamma'$ defined as
\beq
\(\Gamma'\cdot h\)(u)\equiv \oint_{-2g}^{2g}\frac{dv}{{4\pi i}}\d_u \log \frac{\Gamma[i (u-v)+1]}{\Gamma[-i (u-v)]}h(v).
\label{Gammaprime}
\eeq
This requirement is consistent because of the following relation \footnote{We remind that $f_+$ and $f_-$ stand for the part of the Laurent expansion with, respectively, positive and negative powers of $x$, while $\tilde f$ is the analytic continuation around the branch point at $u=2g$ (which amounts to replacing $x\to\ofrac{x}$)}
\beq
\(\Gamma'\cdot h\)(u+i)-\(\Gamma'\cdot h\)(u)
=
-\frac{1}{2\pi i}\oint_{-2g}^{2g}\frac{h(v)}{u-v}dv=h_-(u)-\widetilde{h_+}(u).
\label{eq:Gammaproperty}
\eeq
What is left is to define $\Sigma$ on positive powers of $x$. We do it by requiring
\beq
\frac{1}{2}\Sigma\cdot\left[x^a+1/x^a\right]\equiv p_a'(u) 
\label{paprime}
\eeq
where $p_a'(u)$ is a polynomial in $u$ of degree $a+1$, which is a solution of
\beq
p_a'(u+i)-p_a'(u)=\frac{1}{2}\(x^a+1/x^a\)
\eeq
and satisfies the following additional properties: $p_a'(0)=0$ for odd $a$  and $p_a'(i/2)=0$ for even $a$. One can check that this definition is consistent and defines of $p'_a(u)$ uniquely. Explicit form of the first few $p_a'(u)$, which we call periodized  Chebyshev polynomials, can be found in appendix \ref{sec:appPeriodized}.

From this definition of $\Sigma$ one can see that the result of its action on expressions of the form \eqref{alphabetasinh}
can again be represented in this form - what is important for us is that no exponential functions other than $\sinh(2\pi u)$ appear in the result.

A good illustration of how the definitions above work would be the following two simple examples. Suppose one wants to calculate $\Sigma\cdot\(x-\frac{1}{x}\)$, then it is convenient to split the argument of $\Sigma$ in the following way:
\beq
\Sigma\cdot\(x-\frac{1}{x}\)=\Sigma\cdot\(x+\frac{1}{x}\)-2\Sigma\cdot\frac{1}{x}.
\eeq
In the first term we recognize $p_1'(u)=-\frac{i u(u-i)}{4g}$, whereas in the second the argument of $\Sigma$ is decaying at infinity, thus $\Sigma$ is equivalent to $\Gamma'$ in this context. Notice also that $\Gamma'\cdot \frac{1}{x}=-\Gamma'\cdot x$. All together, we get
\beq
\Sigma\cdot\(x-\frac{1}{x}\)=\Sigma\cdot\(x+\frac{1}{x}\)-2\Sigma\cdot\frac{1}{x}=2p_1'(u)+ 2\Gamma'\cdot x
\eeq

In a similar way, in order to calculate $\Sigma\cdot\frac{\sinh_--\sinh_+}{2}$, one can write $\frac{\sinh_--\sinh_+}{2}=\sinh_--\frac{1}{2}\sinh(2\pi u)$.
Notice that since $\sinh_-$ decays at infinity,
\beq
\Sigma\cdot\sinh_-=\Gamma'\cdot\sinh_-.
\eeq
 Also, since $i$-periodic functions can be factored out of $\Sigma$,
\beq
\Sigma\cdot\sinh(2\pi u)=\sinh(2\pi u)\Sigma\cdot 1=\sinh(2\pi u)p_0'(u)/2.
\eeq
 Finally,
\beq
\Sigma\cdot\frac{\sinh_--\sinh_+}{2}=\Gamma'\cdot(\sinh_-)-\frac{1}{2}\sinh(2\pi u)p_0'(u).
\eeq

As an example we present the particular solution for two components of $\mu_{ab}$ (below we will argue that $\pi_{12}$ and $\pi_{13}$ can be chosen to be zero, see  \eqref{eq:periodicpart})
\beqa
\label{muexpl1}
&&\mu_{13}^{(1)}-\pi_{13}=\Sigma\cdot\({\bf P}_1 \tilde{\bf P}_3-{\bf P}_3 \tilde{\bf P}_1\)=\epsilon^2\Sigma\cdot\(x^2-\frac{1}{x^2}\)=
\epsilon^2\;\;\(\Gamma'\cdot x^2+p_2'(u)\),\\
\nonumber
&&\mu_{12}^{(1)}-\pi_{12}=\Sigma\cdot\({\bf P}_1
   \tilde{\bf P}_2-{\bf P}_2
   \tilde{\bf P}_1\)= \\&& =
   -\epsilon^2\[2 I_1\Gamma'\cdot x-\sinh(2\pi u)\;\Gamma'\cdot x^2-\Gamma'\cdot\(\sinh_-\(x^2+\frac{1}{x^2}\)\)
   \].\label{muexpl2}
\eeqa
Now let us apply $\Sigma$ defined above to \eq{eq:mudiscNLO}, writing that its general solution is
\beq
\mu^{(1)}_{ab}=\Sigma\cdot(\bP_a^{(0)} \tilde\bP_b^{(0)}-  \bP_b^{(0)} \tilde\bP_a^{(0)})+\pi_{ab},
\label{eq:sol13}
\eeq
 where the zero mode $\pi_{ab}$ is an arbitrary $i$-periodic entire function, which can be written similarly to the leading order as $c_{1,ab}\cosh{2\pi u}+c_{2,ab}\sinh{2\pi u}+c_{3,ab}$. Again, many of the coefficients $c_{i,ab}$ can be set to zero. First, the prescription from section \ref{sec:ancont} implies that non-vanishing at infinity part of coefficients of $\sinh(2\pi u)$ and $\cosh(2\pi u)$ in $\mu_{12}$ is zero. As one can see from the explicit form \eqref{muexpl2} of the particular solution which we choose for $\mu_{12}$, it does not contain $\cosh(2\pi u)$ and the coefficient of $\sinh(2\pi u)$ is decaying at infinity. So in order to satisfy the prescription, we have to set $c_{2,12}$ and $c_{2,12}$ to zero. Second, since the coefficients $c_{n,ab}$ are of order $S$, we can remove some of them by making an infinitesimal  $\gamma$-transformation, i.e. with $R=1+{\cal O}(S)$ (see section \ref{sec:Symmetries} and Eq. \eqref{gammaP}).
 Further, the Pfaffian constraint \eqref{constraint} imposes 5 equations on the remaining coefficients, which leaves the following 2-parametric family of zero modes
 \beqa
&& \pi_{12}=0,\ \pi_{13}=0,\ \pi_{14}=\frac{1}{2}c_{1,34}\cosh{2\pi u},\\
 && \pi_{24}= c_{1,24}\cosh{2\pi u},\ \pi_{34}= c_{1,34}\cosh{2\pi u}.
 \eeqa

Let us now look closer at the exponential part of $\mu_{14}$ and $\mu_{24}$. Combining the leading order \eqref{eq:musolLOevenL} and the perturbation \eqref{eq:sol13} and taking into account the fact that operator $\Sigma$ does not produce terms proportional to $\cosh{2 \pi u}$, we obtain
\beqa
&&\mu_{14}=\frac{1}{2}c_{1,34}\cosh{2\pi u}+{\cal O}(\epsilon) \sinh{2\pi u}+\mathcal{O}(\epsilon^2)+\dots, \\
&&\mu_{24}=\frac{1}{2}c_{1,24}\cosh{2\pi u}+(1+{\cal O}(\epsilon)) \sinh{2\pi u}+\mathcal{O}(\epsilon^2)+\dots,
\eeqa
where dots stand for powers-like terms or exponential terms suppressed by powers of $u$.

As we remember from section \ref{sec:ancont}, only the 2nd solution of the 5th order Baxter equation \eqref{5bax} can contribute to the exponential part of $\mu_{14}$ and $\mu_{24}$, which means that $\mu_{14}$ and $\mu_{24}$ are proportional to the same linear combination of $\sinh{2\pi u}$ and $\cosh{2\pi u}$. From the second equation one can see that this linear combination can be normalized to be $\frac{1}{2}c_{1,24}\cosh{2\pi u}+(1+{\cal O}(\epsilon)) \sinh{2\pi u}$. Then $\mu_{14}=C\(\frac{1}{2}c_{1,24}\cosh{2\pi u}+(1+{\cal O}(\epsilon)) \sinh{2\pi u}\)$, where $C$ is some constant, which is of order ${\cal O}(\epsilon)$, because the coefficient of $\sinh{2\pi u}$ in the first equation is ${\cal O}(\epsilon)$. Taking into account that $c_{1,24}$ is ${\cal O}(\epsilon)$ itself, we find that $c_{1,34}=\mathcal{O}(\epsilon^2)$, i.e. it does not contribute at the order which we are considering. So the final form of the zero mode in \eqref{eq:sol13} is
\beqa
&& \pi_{12}=0,\ \pi_{13}=0,\ \pi_{14}=0,\\
 && \pi_{24}=c_{1,24}\cosh{2\pi u},\ \pi_{34}=0.
 \label{eq:periodicpart}
 \eeqa


In this way, using the particular solution given by $\Sigma$ and the form of zero modes \eqref{eq:periodicpart} we have computed all the functions $\mu_{ab}^{(1)}$. The details and the results of the calculation can be found in appendix \ref{sec:appmu2}.


\subsection{Correcting $\bP_{a}$\dots}
\label{sec:CalculationofPa}
In the previous section we found the NLO part of $\mu_{ab}$. Now, according to the iterative procedure described in section \ref{sec:SolvingPmuL2}, we can use it to write a closed system of equations for $\bP_a^{(1)}$.
Indeed, expanding the system \eqref{eq:pmuexpanded} to NLO we get
\beqa
\label{eq:P1eqNLOL2}
&&\tilde \bP^{(1)}_1
- \bP^{(1)}_1
= -\bP^{(1)}_3+r_1,  \\
\label{eq:P2eqNLOL2}
&&\tilde \bP^{(1)}_2+\bP_2^{(1)}= -\bP^{(1)}_4  -\bP^{(1)}_1 \sinh(2\pi u)+r_2, \\
\label{eq:P3eqNLOL2}
&&\tilde \bP^{(1)}_3+\bP_3^{(1)}=r_3,\\
\label{eq:P4eqNLOL2}
&&\tilde \bP^{(1)}_4-\bP_4^{(1)}=\bP_3^{(1)} \sinh(2\pi u)+r_4,
\eeqa
where the free terms are given by
\beq
r_a=-\mu_{ab}^{(1)}\chi^{bc}\bP_c^{(0)}.
\label{eq:ra}
\eeq
Notice that $r_a$ does not change if we add a matrix proportional to $\bP_a^{(0)} \tilde\bP_b^{(0)}-  \bP_b^{(0)} \tilde\bP_a^{(0)}$ to $\mu^{(1)}_{ab}$, due to the relations
\beq
	\bP_a \chi^{ab}\bP_b=0,\;\bP_a\chi^{ab}\tilde\bP_b=0,
\eeq	
which follow from the $\bP\mu$-system equations. In particular we can use this property to replace $\mu_{ab}^{(1)}$ in \eq{eq:ra} by $\mu_{ab}^{(1)}+\frac{1}{2}\(\bP_a^{(0)} \tilde\bP_b^{(0)}-  \bP_b^{(0)} \tilde\bP_a^{(0)}\)$. This will be convenient for us, since in expressions for $\mu^{(1)}_{ab}$ in terms of $p_a$ and $\Gamma$ (see \eq{muexpl1}, \eq{muexpl2} and appendix \ref{sec:appmu2}) this change amounts to simply replacing $\Gamma'$ by a convolution with a more symmetric kernel:
\beq
 \Gamma' \rightarrow  \Gamma,
\eeq
\beq
\(\Gamma\cdot h\)(u)\equiv \oint_{-2g}^{2g}\frac{dv}{{4\pi i}}\d_u \log \frac{\Gamma[i (u-v)+1]}{\Gamma[-i (u-v)+1]}h(v),
\label{Gamma}
\eeq
while at the same time replacing
\beq
	 p_a'(u)\rightarrow  p_a(u),\ \
\eeq
\beq
	p_a(u)=p_a'(u)+\frac{1}{2}\(x^a(u)+x^{-a}(u)\).
\label{pa}
\eeq

Having made this comment, we will now develop tools for solving the equations \eq{eq:P1eqNLOL2} - \eq{eq:P4eqNLOL2}.
Notice first that if we solve them in the order \eq{eq:P3eqNLOL2}, \eq{eq:P1eqNLOL2}, \eq{eq:P4eqNLOL2}, \eq{eq:P2eqNLOL2}, substituting into each subsequent equation the solution of all the previous, then at each step the problem we have to solve has a form

\beq
 \tilde f+f=h\;\; \text{or}\;\; \tilde f-f=h\;\;,
 \label{eqs}
\eeq
  where $h$ is known, $f$ is unknown and both the right hand side and the left hand side are power series in $x$. It is obvious that equations \eqref{eqs} have solutions only for $h$ such that $h=\tilde h$ and $h=-\tilde h$ respectively.
   On the class of such $h$ a particular solution for $f$ can be written as
  \beq
  f= [h]_-+[h]_0/2\equiv H\cdot h\;\; \Rightarrow\;\; \tilde f+f=h
    \label{solfh1}
  \eeq
  and
    \beq
    f= -[h]_-\equiv K\cdot h\;\; \Rightarrow\;\; \tilde f-f=h,
  \label{solfh2}
  \eeq
  where $[h]_0$ is the constant part of Laurent expansion of $h$ (it does not appear in the second equation, because $h$ such that $h=-\tilde h$ does not have a constant part).
  The operators $K$ and $H$ introduced here can be also defined by their integral kernels
\beqa
H(u,v)&=&-\frac{1}{4\pi i}\frac{\sqrt{u-2g}\sqrt{u+2g}}{\sqrt{v-2g}\sqrt{v+2g}}\frac{1}{u-v}dv, \\
K(u,v)&=&+\frac{1}{4\pi i}\frac{1}{u-v}dv.
\label{eq:HK}
\eeqa
which are equivalent to \eqref{solfh1},\eqref{solfh2} of the classes of $h$ such that $h=\tilde h$ and $h=-\tilde h$ respectively\footnote{We denote e.g. $K\cdot h=\oint_{-2g}^{2g}K(u,v)h(v)dv$ where the integral is around the branch cut between $-2g$ and $2g$.}. The particular solution $f=H\cdot h$ of the equation $\tilde f+ f=h$ is unique in the class of functions $f$ decaying at infinity, and the solution $f=K \cdot h$ of $\tilde f- f=h$ is unique for non-growing $f$. In all other cases the general solution will include zero modes, which, in our case are fixed by asymptotics of $\bP_a$.

Now it is easy to write the explicit solution of the equations
\eq{eq:P1eqNLOL2}-\eq{eq:P4eqNLOL2}:
\beqa
\bP_3^{(1)}&=&H\cdot r_3,\\
\bP_1^{(1)}&=&\frac{1}{2}\P^{(1)}_3+K\cdot \(r_1-\frac{1}{2} r_3\),\\
\bP_4^{(1)}&=&K\cdot\(-\frac{1}{2}\(\tilde\bP_3^{(1)}-\bP_3^{(1)}\) \sinh(2\pi u)+
\frac{2r_4+r_3 \sinh(2\pi u)}{2}\)-2\delta,\\
\bP_2^{(1)}&=&H\cdot\(-\frac{1}{2}
\(
{\bf P}^{(1)}_4+\sinh(2\pi u){\bf P}^{(1)}_1+\tilde{\bf P}^{(1)}_4+\sinh(2\pi u)\tilde{\bf P}^{(1)}_1
\)+\right.\\ \nn
&&\left.
+\frac{r_4+\sinh(2\pi u) r_1+2r_2}{2}\)+\delta,
\label{eq:P4solNLOL2}
\eeqa
where $\delta$ is a constant fixed uniquely by requiring $\mathcal{O}(1/u^2)$ asymptotics for $\bP_2$. This asymptotic also sets the last coefficient $c_{1,24}$ left in $\pi_{12}$ to zero. Thus in the class of functions with asymptotics \eqref{eq:asymptotics} the solution for $\mu_{ab}$ and $\bP_a$ is unique up to a $\gamma$-transformation.

\subsection{Result for $J=2$}
\label{sec:resultL2}

In order to obtain the result for the anomalous dimension, we again use the formulas \eq{AA1}, \eq{AA2} which connect the leading coefficients of $\bP_a$ with $\Delta,\ J\ $ and $S$. After plugging in $A_i$ which we find from our solution, we obtain the result for the $S^2$ correction to the anomalous dimension:
\beqa
\label{gamma2L2}
\gamma^{(2)}_{J=2}&=&\frac{\pi}{g^2(I_1-I_3)^3}\oint \frac{du_x}{2\pi i}\oint \frac{du_y}{2\pi i}\[\frac{8  I_1^2(I_1+I_3) \left(x^3-\left(x^2+1\right) y\right) }{ \left(x^3-x\right) y^2}\right.\\ \nn
&&   +\frac{8  \sh_-^x \sh_-^y
   \left(x^2 y^2-1\right) \left(I_1 (x^4 y^2+1)-I_3x^2(y^2+1)\right)}{ x^2 \left(x^2-1\right)
   y^2}\\ \nn
&&-\frac{4  (\sh_-^y)^2 x^2 \left(y^4-1\right) \left( I_1(2x^2-1)-I_3 \right)}{ \left(x^2-1\right) y^2}\\ \nn
&&+\frac{8
   I_1^2 \sh_-^y x  \left(2 \(x^3-x\) \left(y^3+y\right)-2 x^2
   \left(y^4+y^2+1\right)+y^4+4 y^2+1\right)}{ \left(x^2-1\right) y^2}\\  \nn
&&-\frac{8 (I_1-I_3)
   I_1 \sh_-^y x   (x-y) (x
   y-1)}{ \left(x^2-1\right) y}\\ \nn
&&\left.-\frac{4 (I_1-I_3) (\sh_-^x)^2 \left(x^2+1\right)
   y^2}{ \left(x^2-1\right)}\right]
	\frac{1}{4\pi i}\d_u \log\frac{\Gamma (i u_x-i u_y+1)}{\Gamma (1-i u_x+i u_y)}\;.
\eeqa
Here the integration contour goes around the branch cut at $(-2g,2g)$. We also denote
$\sh_-^x=\sinh_-(x) ,\ \sh_-^y=\sinh_-(y) $ (recall that $\sinh_-$ was defined in \eq{defshm}). This is our final result for the curvature function at any coupling.

It is interesting to note that our result contains the combination $\log\frac{\Gamma (i u_x-i u_y+1)}{\Gamma (1-i u_x+i u_y)}$ which plays an essential role in the construction of the BES dressing phase. We will use this identification in section \ref{sec:Konishidimension} to compute the integral in \eq{gamma2L2} numerically with high precision.

In the next subsections we will describe generalizations of the $J=2$ result to operators with $J=3$ and $J=4$.

 \subsection{Results for higher $J$}
\label{sec:SolvingPmuL3}

Solving the $\bP\mu$-system for $J=3$ is similar to the $J=2$ case described above, except for several technical complications, which we will describe here, leaving the details for the appendix \ref{sec:appnlo3}.
As in the previous section, the starting point is the LO solution of the $\bP\mu$ system, which for $J=3$ reads
\beq
	\bP_1=\epsilon x^{-3/2},\ \bP_3=-\epsilon x^{3/2},
\label{P1P3LOsolL3}
\eeq
\beq
	\bP_2=-\epsilon x^{3/2}\cosh_- +\epsilon x^{-1/2}I_2,
\label{P2LOsolL3}
\eeq
\beq
	\bP_4=-\epsilon x^{1/2}I_2-\epsilon x^{-3/2}I_0-\epsilon x^{-3/2}\cosh_-,
\label{P4LOsolL3}
\eeq
\beq
	\mu_{12}=1,\ \mu_{13}=0,\ \mu_{14}=0,\  \mu_{24}=\cosh(2\pi u),\ \mu_{34}=1\;.
\eeq
The first step is to construct $\mu^{(1)}_{ab}$ from its discontinuity given by the equation \eqref{eq:mudiscNLO}. The full solution consists of a particular solution and a general solution of the corresponding homogeneous equation, i.e. zero mode $\pi_{ab}$. In our case the zero mode can be an $i$-periodic function, i.e. a linear combination of $\sinh(2\pi u)$, $\cosh(2\pi u)$ and constants. As in the case of $J=2$, we use a combination of the Pfaffian constraint, prescription from section \ref{sec:ancont} and a $\gamma$-transformation to reduce all the parameters of the zero mode to just one, sitting in $\mu_{24}$:
 \beq
\pi_{12}=0,\;\pi_{13}=0,\;\pi_{14}=0,\;\pi_{24}=c_{24,2} \sinh\(2\pi u\),\;\pi_{34}=0.
\label{eq:periodicpartL3}
\eeq

As in the previous section, the next step is to find $\bP_a^{(1)}$ from the $P\mu$ system expanded to the first order, namely from
\beqa
\label{P1L3}
&&\tilde \bP_1^{(1)}+\bP_3^{(1)}=r_1,\\
&&\tilde \bP_2^{(1)}+\bP_4^{(1)}+\bP_1^{(1)} \cosh(2\pi u)=r_2,\\
&&\tilde \bP_3^{(1)}+\bP_1^{(1)}=r_3,\\
&&\tilde \bP_4^{(1)}+\bP_2^{(1)}-\bP_3^{(1)}\cosh(2\pi u)=r_4,
\label{P4L3}
\eeqa
where $r_a$ are defined by \eqref{eq:ra} and for $J=3$ are given explicitly in appendix \ref{sec:appnlo3}.
In attempt to solve this system, however, we encounter another technical complication. As one can see from \eqref{P1P3LOsolL3}-\eqref{P4LOsolL3}, the LO solution contains half-integer powers of $J$, meaning that the $\bP_a$ now have an extra branch point at infinity.
However, the operations $H$ and $K$ defined by \eq{eq:HK} work only for functions which have Laurent expansion in integer powers of $x$. In order to solve equations of the type \eq{eq:mudiscNLO} on the class of functions which allow Laurent-like expansion in $x$ with only half-integer powers $x$, we introduce operations $H^*,K^*$:
\beqa
&&H^*\cdot f\equiv\frac{x+1}{\sqrt{x}}H\cdot\frac{\sqrt{x}}{x+1} f, \\
&&K^*\cdot f\equiv\frac{x+1}{\sqrt{x}}K\cdot\frac{\sqrt{x}}{x+1} f.
\eeqa
In terms of these operations the solution of the system \eqref{P1L3}-\eqref{P4L3} is
\beqa
\label{P1J3}
\bP_{1}^{(1)}&=&\frac{1}{2}\(H^*(r_1+r_3)+ K^*(r_1-r_3)\)+\bP_1^{\text{zm}},\\
\bP_{3}^{(1)}&=&\frac{1}{2}\(H^*(r_1+r_3)- K^*(r_1-r_3)\)+\bP_2^{\text{zm}},\\
\bP_{2}^{(1)}&=&\frac{1}{2}\(H^*(r_2+r_4)+ K^*(r_2-r_4)-\right.\nonumber\\
&-&\left.H^*\(\cosh(2\pi u)K^*(r_1-r_3)\)- K^*\(\cosh(2\pi u)H^*(r_1+r_3)\)\right)+\bP_3^{\text{zm}},\\
\label{P4J3}
\bP_{4}^{(1)}&=&\frac{1}{2}\(H^*(r_2+r_4)- K^*(r_2-r_4)-\right.\nonumber\\
&-&\left.H^*\(\cosh(2\pi u)K^*(r_1-r_3)\)+ K^*\(\cosh(2\pi u)H^*(r_1+r_3)\)\right)+\bP_4^{\text{zm}},
\eeqa
where $\bP_a^{\text{zm}}$ is a solution of the system \eq{P1L3}-\eq{P4L3} with right hand side set to zero, whose explicit form $\bP_a^{\text{zm}}$ is given in Appendix \ref{sec:appnlo3} (see \eqref{P1J3zm}-\eqref{P4J3zm}) and which is parametrized by four constants $L_1,L_2,L_3,L_4$, e.g.
\beqa
\bP_1^{\text{zm}}=L_1 x^{-1/2}+L_3x^{1/2}.
\eeqa
These constants are fixed by requiring correct asymptotics of $\bP_a$, which also fixes the parameter $c_{24,2}$ in the zero mode \eq{eq:periodicpartL3} of $\mu_{ab}$ \footnote{Actually in this way $c_{24,2}$ is fixed to be zero.}. Indeed, a priori $\bP_2$ and $\bP_1$ have wrong asymptotics. Imposing a constraint that $\bP_2$ decays as $u^{-5/2}$ and $\bP_1$ decays as $u^{-3/2}$ produces five equations, which fix all the parameters uniquely.

Skipping the details of the intermediate calculations, we present the final result for the anomalous dimension:
\beqa
\label{gamma2L3}
&&\gamma^{(2)}_{J=3}=\oint \frac{du_x}{2\pi i}\oint \frac{du_y}{2\pi i}
i \frac{1}{g^2(I_2-I_4)^3} \left[\frac{2 \left(x^6-1\right) y (\ch_-^y)^2 (I_2-I_4)}{x^3 \left(y^2-1\right)}-\right.\\ \nn
&&-\frac{4 \ch_-^x
   \ch_-^y \left(x^3 y^3-1\right) \left(I_2 x^5 y^3+I_2-I_4 x^2 \left(x y^3+1\right)\right)}{x^3
   \left(x^2-1\right) y^3}+\\ \nn
&& +\frac{(y^2-1)  (\ch_-^y)^2 I_2 \left( (x^8+1) \left(2 y^4+3 y^2+2\right)-(x^6+x^2)
   \left(y^2+1\right)^2\right)}{x^3 \left(x^2-1\right) y^3}-
   \\ \nn
   && -\frac{(y^2-1)  (\ch_-^y)^2 I_4 \left((x^8+1) y^2+(x^6+x^2) \left(y^4+1\right)\right)}{x^3 \left(x^2-1\right) y^3}-
   \\ \nn
&&-\frac{4 I_2 \ch_-^y (x-y) (x y-1) \left(I_2
   \left(\(x^6+1\) \left(y^3+y\right)+\(x^5+x\) \left(y^4+y^2+1\right)-x^3 \left(y^4+1\right)\right)+I_4 x^3
   y^2\right)}{x^3 \left(x^2-1\right) y^3}-\\ \nn
&& \left.-\frac{I_2^2 (y^2-1)  (x-y) (x y-1) \left(I_2 \left(\(x^6 +x^4 +x^2 +1\)y+2 x^3
   \left(y^2+1\right)\right)+I_4 \left(x^5+x\right) \left(y^2+1\right)\right)}{x^3 \left(x^2-1\right) y^3}\right]\\ \nn
	&& \frac{1}{4\pi i}\d_u \log\frac{\Gamma (i u_x-i u_y+1)}{\Gamma (1-i u_x+i u_y)}.
\eeqa
We defined $\ch_-^x=\cosh_-(x)$ and $\ch_-^y=\cosh_-(y)$, where $\cosh_-(x)$ is the part of the Laurent expansion of $\cosh\(g(x+1/x)\)$ vanishing at infinity, i.e.
\beq
\cosh_-(x)=\sum_{k=1}^{\infty} I_{2k}x^{-2k}.
\eeq
The result for $J=4$ is given in appendix \ref{sec:SolvingPmuL4}.

\section{Weak coupling tests and predictions}
\label{sec:weak}

Our results for the curvature function $\gamma^{(2)}(g)$ at $J=2,3,4$ (Eqs. \eq{gamma2L2}, \eq{gamma2L3}, \eq{gamma2L4}) are straightforward to expand at weak coupling. We give expansions to 10 loops in \text{appendix \ref{sec:weakS3}}. Let us start with the $J=2$ case, for which we found
\beqa
\label{weak22}
\gamma_{J=2}^{(2)}&=&-8 g^2 \zeta_3+g^4 \left(140 \zeta_5-\frac{32 \pi ^2 \zeta_3}{3}\right)+g^6 \left(200 \pi ^2 \zeta_5-2016
   \zeta_7\right)
	\\ \nn
	&+&g^8 \left(-\frac{16 \pi ^6 \zeta_3}{45}-\frac{88 \pi ^4 \zeta_5}{9}-\frac{9296 \pi ^2 \zeta_7}{3}+27720 \zeta_9\right)
	\\ \nn
	&+&g^{10} \left(\frac{208 \pi ^8 \zeta_3}{405}+\frac{160 \pi ^6 \zeta_5}{27}+144
   \pi ^4 \zeta_7+45440 \pi ^2 \zeta_9-377520 \zeta_{11}\right)
	+\dots
\eeqa
Remarkably, at each loop order all contributions have the same transcendentality, and only simple zeta values (i.e. $\zeta_n$) appear. This is also true for the $J=3$ and $J=4$ cases.

We can check this expansion against known results, as the anomalous dimensions of twist two operators have been computed up to five loops for arbitrary spin \cite{Kotikov:2001sc,Kotikov:2003fb,Kotikov:2004er,Moch:2004pa,Staudacher:2004tk,Kotikov:2007cy,Bajnok:2008qj,Lukowski:2009ce} (see also \cite{Velizhanin:2013vla} and the review \cite{Freyhult:2010kc}).
To three loops they can be found solely from the ABA equations, while at four and five loops wrapping corrections need to be taken into account which was done in \cite{Bajnok:2008qj,Lukowski:2009ce} by utilizing generalized Luscher formulas. All these results are given by linear combinations of harmonic sums
\beq
	S_a(N) = \sum_{n=1}^N\frac{(\mathrm{sign}(a))^n}{n^{|a|}}, \ \
	S_{a_1,a_2,a_3,\dots}(N)=\sum_{n=1}^N\frac{(\mathrm{sign}(a_1))^n}{n^{|a_1|}}S_{a_2,a_3,\dots}(n)
\eeq
with argument equal to the spin $S$. To make a comparison with our results we expanded these predictions in the $S\to 0$ limit. For this lengthy computation, as well as to simplify the final expressions, we used the \verb"Mathematica" packages HPL \cite{HPL}, the package \cite{VolinPackage} provided with the paper \cite{Leurent:2013mr}, and the HarmonicSums package \cite{Ablinger}.

In this way we have confirmed the coefficients in \eq{weak22} to four loops. Let us note that expansion of harmonic sums leads to multiple zeta values (MZVs), which however cancel in the final result leaving only $\zeta_n$.

Importantly, the part of the four-loop coefficient which comes from the wrapping correction is essential for matching with our result. This is a strong confirmation that our calculation based on the $\bP\mu$-system is valid beyond the ABA level. Additional evidence that our result incorporates all finite-size effects is found at strong coupling (see section \ref{sec:SlopeSlopeStrongCoupling}).

For operators with $J=3$, our prediction at weak coupling is
\beqa
	\gamma_{J=3}^{(2)}&=&-2g^2\zeta_3+g^4\(12 \zeta_5-\frac{4 \pi ^2 \zeta_3}{3}\)
	+g^6\(\frac{2 \pi ^4 \zeta_3}{45}+8 \pi ^2 \zeta_5-28
   \zeta_7\)\\ \nn
   &+&
   g^8\(-\frac{4 \pi ^6 \zeta_3}{45}-\frac{4 \pi ^4 \zeta_5}{15}-528 \zeta_9\)
   +\dots
\eeqa
The known results for any spin in this case are available at up to six loops, including the wrapping correction which first appears at five loops \cite{Beccaria:2007cn,Beccaria:2009eq,Velizhanin:2010cm}. Expanding them at $S\to 0$ we have checked our calculation to four loops.\footnote{As a further check it would be interesting to expand to order $S^2$ the known results for twist 2 operators at five loops, and for twist 3 operators at five and six loops -- all of which are given by huge expressions.}

For future reference, in appendix \ref{sec:weakS3} we present an expansion of known results for $J=2,3$ up to order $S^3$ at first several loop orders. In particular, we found that multiple zeta values appear in this expansion, which did not happen at lower orders in $S$.

\FIGURE[ht]
{
\label{fig:j4plot}

    \begin{tabular}{cc}
    \includegraphics[scale=0.9]{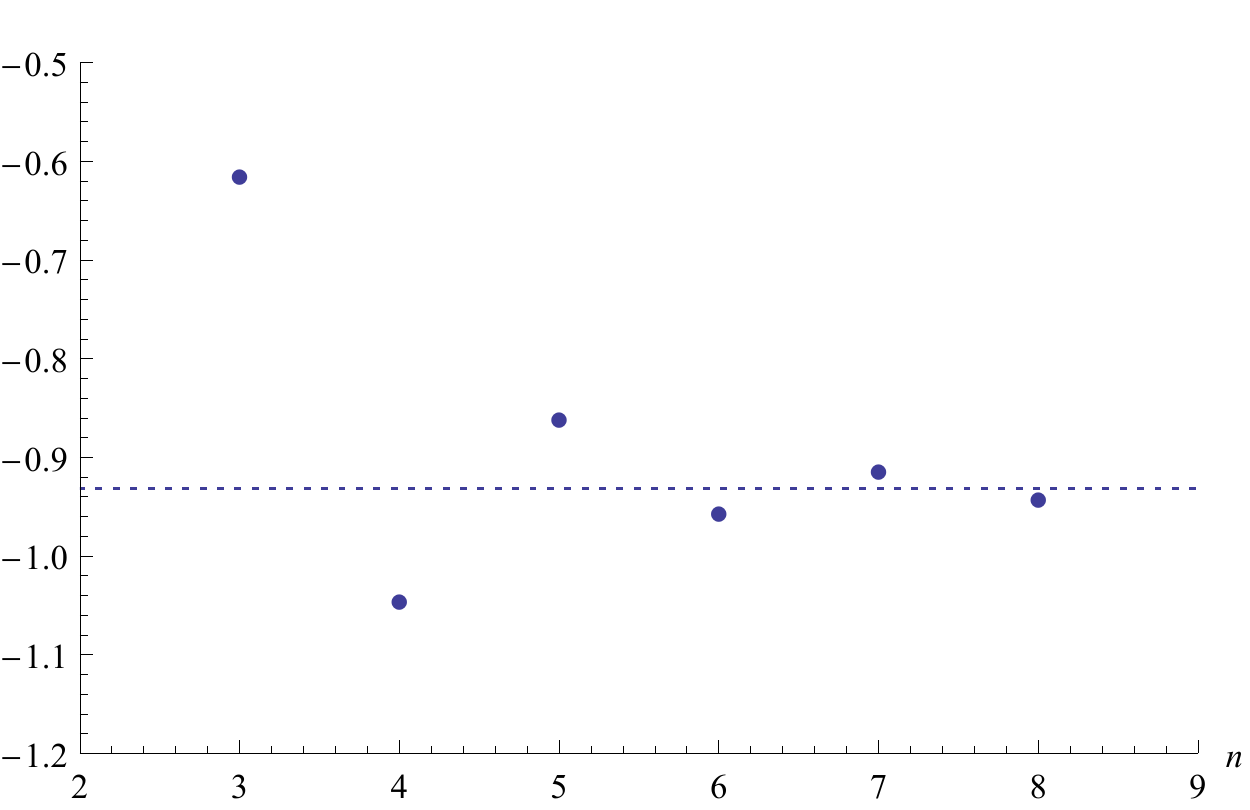}\\
    \end{tabular}
\caption{\textbf{One-loop energy at $J=4$ from the Bethe ansatz.} The dashed line shows the result from the $\bP\mu$-system for the coefficient of $S^2$ in the 1-loop energy at $J=4$, i.e. $-\frac{14 \zeta_3}{5}+\frac{48 \zeta_5}{\pi ^2}-\frac{252 \zeta_7}{\pi
   ^4}\approx-0.931$ (see \eq{gamma42weak}). The dots show the Bethe ansatz prediction \eq{gammaJ} expanded to orders $1/J^3,1/J^4,\dots,1/J^8$ (the order of expansion $n$ corresponds to the horizontal axis), and it appears to converge to the $\bP\mu$-system result.
}
}

Let us now discuss the $J=4$ case. The expansion of our result reads:
\beqa
\label{gamma42weak}
	\gamma_{J=4}^{(2)}&=&g^2 \left(-\frac{14 \zeta_3}{5}+\frac{48 \zeta_5}{\pi ^2}-\frac{252 \zeta_7}{\pi
   ^4}\right)	\\ \nn
	&+&g^4 \left(-\frac{22 \pi ^2 \zeta_3}{25}+\frac{474 \zeta_5}{5}-\frac{8568 \zeta_7}{5 \pi
   ^2}+\frac{8316 \zeta_9}{\pi ^4}\right)\\ \nn
	&+&g^6 \left(\frac{32 \pi ^4 \zeta_3}{875}+\frac{3656 \pi ^2 \zeta_5}{175}-\frac{56568 \zeta_7}{25}+\frac{196128 \zeta_9}{5 \pi ^2}-\frac{185328 \zeta_{11}}{\pi ^4}\right)	\\ \nn
	&+&g^8 \left(-\frac{4 \pi ^6 \zeta_3}{175}-\frac{68 \pi ^4 \zeta_5}{75}-\frac{55312 \pi ^2 \zeta_7}{125}+\frac{1113396 \zeta_9}{25}-\frac{3763188 \zeta_{11}}{5 \pi ^2}\right.
	\\ \nn
	 && \ \ \ \ \ \  + \left.\frac{3513510 \zeta_{13}}{\pi ^4} \right)+\dots
\eeqa
Unlike for the $J=2$ and $J=3$ cases, we could not find a closed expression for the energy at any spin $S$ in literature even at one loop, however there is another way to check our result. One can expand the asymptotic Bethe ansatz equations at large $J$ for fixed values of $S=2,4,6,\dots$ and then extract the coefficients in the expansion which are polynomial in $S$. This was done in \cite{Beccaria:2012kp} (see appendix C there) where at one loop the expansion was found up to order $1/J^6$:
\beqa
\label{gammaJ}
\nn
	\gamma(S,J) &=& g^2\(\frac{S}{2\,J^{2}}-\Big(\frac{S^{2}}{4}+\frac{S}{2}\Big)\,\frac{1}{ J^{3}}
+\Big[
\frac{3 S^3}{16}+\Big(\frac{1}{8}-\frac{\pi ^2}{12}\Big) S^2+\frac{S}{2}
\Big]\,\frac{1}{ J^{4}} +\dots\)+\mathcal{O}(g^4)
\\
\eeqa
Now taking the part proportional to $S^2$ and substituting $J=4$ one may expect to get a numerical approximation to the 1-loop coefficient in our result \eq{gamma42weak}, i.e. $-\frac{14 \zeta_3}{5}+\frac{48 \zeta_5}{\pi ^2}-\frac{252 \zeta_7}{\pi
   ^4}$. To increase the precision we extended the expansion in \eq{gammaJ} to order $1/J^8$. Remarkably, in this way we confirmed the 1-loop part of the $\bP\mu$ prediction \eq{gamma42weak} with about $1\%$ accuracy! In Fig. \ref{fig:j4plot} one can also see that the ABA result converges to our prediction when the order of expansion in $1/J$ is being increased.

Also, in contrast to $J=2$ and $J=3$ cases we see that negative powers of $\pi$ appear in \eq{gamma42weak} (although still all the contributions at a given loop order have the same transcendentality). It would be interesting to understand why this happens  from the gauge theory perspective, especially since expansion of the leading $S$ term \eq{slopeIn} has the same structure for all $J$,
\beq
	\gamma_{J}^{(1)}=\frac{8 \pi ^2 g^2}{J (J+1)}-\frac{32 \pi ^4 g^4}{J (J+1)^2
   (J+2)}+\frac{256 \pi ^6 g^6}{J (J+1)^3 (J+2) (J+3)}
	+\dots
\eeq
The change of structure at $J=4$ might be related to the fact that for $J\geq 4$ the ground state anomalous dimension even at one loop is expected to be an irrational number for integer $S>0$ (see \cite{Beccaria:2008pp}, \cite{Belitsky:2008mg}), and thus cannot be written as a linear combination of harmonic sums with integer coefficients.

In the next section we will discuss tests and applications of our results at strong coupling.

\section{Strong coupling tests and predictions}
\label{sec:SlopeSlopeStrongCoupling}

In this section we will present the strong coupling expansion of our results for the curvature function, and link these results to anomalous dimensions of short operators at strong coupling. We will also obtain new predictions for the BFKL pomeron intercept.

\subsection{Expansion of the curvature function for $J=2,3,4$}

To obtain the strong coupling expansion of our exact results for the curvature function, we evaluated it numerically with high precision for a range of values of $g$ and then made a fit to find the expansion coefficients. It would also be interesting to carry out the expansion analytically, and we leave this for the future.

For numerical study it is convenient to write our exact expressions \eq{gamma2L2}, \eq{gamma2L3}, \eq{gamma2L4} for $\gamma^{(2)}(g)$, which have the form
\beq
\label{ints}
	\gamma^{(2)}(g)=\oint {du_x}\oint {du_y} f(x,y) \d_{u_x} \log\frac{\Gamma (i u_x-i u_y+1)}{\Gamma (1-i u_x+i u_y)}
\eeq
where the integration goes around the branch cut between $-2g$ and $2g$, in a slightly different way (we remind that we use notation $x+\ofrac{x}=\frac{u_x}{g}$ and $y+\ofrac{y}=\frac{u_y}{g}$). Namely, by changing the variables of integration to $x,y$ and integrating by parts one can write the result as
\beq
\label{ints2}
	\gamma^{(2)}(g)=\oint {dx}\oint {dy} F(x,y) \log\frac{\Gamma (i u_x-i u_y+1)}{\Gamma (i u_y-i u_x+1)}
\eeq
where $F(x,y)$ is some polynomial in the following variables: $x,\;1/x,\;y,\;1/y,\;\sh_-^x$ and $\sh_-^y$ (for $J=3$ it includes $\ch_-^x,\;\ch_-^y$ instead of the $\sh_-$ functions). The integral in \eq{ints2} is over the unit circle. The advantage of this representation is that plugging in $\sh_-^x$, $\sh_-^y$ as series expansions (truncated to some large order), we see that it only remains to compute integrals of the kind
\beqa
C_{r,s}&=&\frac{1}{i}\oint\frac{dx}{2\pi}\oint\frac{dy}{2\pi}x^r y^s\log\frac{\Gamma(i u_x-iu_y+1)}{\Gamma(i u_y-iu_x+1)}
\eeqa
These are nothing but the coefficients of the BES dressing phase \cite{Beisert:2006ez,Dorey:2007xn,Beisert:2006ib,Vieira:2010kb}. They can be conveniently computed using the strong coupling expansion \cite{Beisert:2006ez}
\beq
C_{r,s}=\sum_{n=0}^\infty\[-\frac{2^{-n-1} (-\pi )^{-n} g^{1-n} \zeta_n
   \left(1-(-1)^{r+s+4}\right) \Gamma \left(\frac{1}{2}
   (n-r+s-1)\right) \Gamma \left(\frac{1}{2}
   (n+r+s+1)\right)}{\Gamma (n-1) \Gamma \left(\frac{1}{2}
   (-n-r+s+3)\right) \Gamma \left(\frac{1}{2} (-n+r+s+5)\right)}\]
\eeq
However this expansion is only asymptotic and does not converge. For fixed $g$ the terms will start growing with $n$ when $n$ is greater than some value $N$, and we only summed the terms up to $n=N$ which gives the value of $C_{r,s}$ with very good precision for large \text{enough $g$}.

Using this approach we computed the curvature function for a range of values of $g$ (typically we took $7\leq g \leq 30$) and then fitted the result as an expansion in $1/g$. This gave us only numerical values of the expansion coefficients, but in fact we found that with very high precision the coefficients are as follows. For $J=2$
\beqa
\label{eq:ssj2}
\gamma^{(2)}_{J=2}&=&-\pi ^2 g^2+\frac{\pi  g}{4}+\frac{1}{8}-\ofrac{\pi g}\(\frac{3 \zeta_3}{16}+\frac{3}{512}\)-\frac{1}{\pi^2g^2}\(\frac{9 \zeta_3}{128}+\frac{21}{512}\)
\\ \nn
&+&
\frac{1}{\pi^3g^3}\(\frac{3 \zeta_3}{2048}+\frac{15 \zeta_5}{512}-\frac{3957}{131072}\) + \dots\;,
\eeqa
then for $J=3$
\beqa
\label{eq:ssj3}
\gamma^{(2)}_{J=3}&=&-\frac{8 \pi ^2 g^2}{27}+\frac{2 \pi  g}{27}+\frac{1}{12}-
\ofrac{\pi g}\(
\frac{1}{216}
+\frac{\zeta_3}{8}
\)-
\frac{1}{\pi^2g^2}\(\frac{3 \zeta_3}{64}+\frac{743}{13824}\)
\\ \nn
&+&
\frac{1}{\pi^3g^3}\(\frac{41 \zeta_3}{1024}+\frac{35 \zeta_5}{512}-\frac{5519}{147456}\) + \dots\;,
\eeqa
and finally for $J=4$
\beqa
\gamma^{(2)}_{J=4}&=&-\frac{\pi ^2 g^2}{8}+\frac{\pi  g}{32}+\frac{1}{16}-\ofrac{\pi g}\(\frac{3 \zeta_3}{32}+\frac{15}{4096}\)-\frac{0.01114622551913}{g^2}
\\ \nn
&+&\frac{0.004697583899}{g^3}+ \dots\;.
\eeqa
To fix coefficients for the first four terms in the expansion we were guided by known analytic predictions which will be discussed below, and found that our numerical result matches these predictions with high precision. Then for $J=2$ and $J=3$ we extracted the numerical values obtained from the fit for the coefficients of $1/g^2$ and $1/g^3$, and plugging them into the online calculator EZFace \cite{ezface} we obtained a prediction for their exact values as combinations of $\zeta_3$ and $\zeta_5$. Fitting again our numerical results with these exact values fixed, we found that the precision of the fit at the previous orders in $1/g$ increased. This is a highly nontrivial test for the proposed exact values of $1/g^2$ and $1/g^3$ terms. For $J=2$ we confirmed the coefficients of these terms with absolute precision $10^{-17}$ and $10^{-15}$ at $1/g^2$ and $1/g^3$ respectively (at previous orders of the expansion the precision is even higher). For $J=3$ the precision was correspondingly $10^{-15}$ and $10^{-13}$.

For $J=4$ we were not able to get a stable fit for the $1/g^2$ and $1/g^3$ coefficients from EZFace, so above we gave their numerical values (with uncertainty in the last digit). However below we will see that based on $J=2$ and $J=3$ results one can make a prediction for these coefficients, which we again confirmed by checking that precision of the fit at the previous orders in $1/g$ increases. The precision of the final fit at orders $1/g^2$ and $1/g^3$ is $10^{-16}$ and $10^{-14}$ respectively.

\subsection{Generalization to any $J$}
Here we will find an analytic expression for the strong coupling expansion of the curvature function which generalizes the formulas \eqref{eq:ssj2} and \eqref{eq:ssj3} to any $J$. To this end it will be beneficial to consider the structure of classical expansions of the scaling dimension. A good entry point is considering the inverse relation $S(\Delta)$, frequently encountered in the context of BFKL. It satisfies a few basic properties, namely the curve $S(\Delta)$ goes through the points $(\pm J, 0)$ at any coupling, because at $S=0$ the operator is BPS. At the same time for non-BPS states
one should have $\Delta(\lambda)\propto \lambda^{1/4}\to\infty$ \cite{Gubser:1998bc} which indicates that if $\Delta$ is fixed, $S$ should go to zero, thus combining this with the knowledge of fixed points $(\pm J, 0)$ we conclude that at infinite coupling $S(\Delta)$ is simply the line $S = 0$. As the coupling becomes finite $S(\Delta)$ starts bending from the $S=0$ line and starts looking like a parabola going through the points $\pm J$, see fig. \ref{pic:bfkl}. Based on this qualitative picture and the scaling $\Delta(\lambda)\propto \lambda^{1/4}$ at $\lambda\rightarrow\infty$ and fixed $J$ and $S$, one can write down the following ansatz,
\beqa
\label{eq:sofdelta}
	S(\Delta) &=& \left( \Delta^2 - J^2\right)\Bigl( \alpha_1 \frac{1}{\lambda^{1/2}} + \alpha_2 \frac{1}{\lambda} + (\alpha_3 + \beta_3 \Delta^2) \frac{1}{\lambda^{3/2}} + (\alpha_4 + \beta_4 \Delta^2) \frac{1}{\lambda^{2}}   \Bigr.\\
	\nn
	&+&
	\Bigl.
	(\alpha_5 + \beta_5 \Delta^2+\gamma_5\Delta^4) \frac{1}{\lambda^{5/2}}
	+(\alpha_6 + \beta_6 \Delta^2+\gamma_6\Delta^4) \frac{1}{\lambda^{3}}
	+\dots
	\Bigr).
\eeqa
We omit odd powers of the scaling dimension from the ansatz, as only the square of $\Delta$ enters the $\bf{P}\mu$-system. We can now invert the relation and express $\Delta$ in terms of $S$ at strong coupling, which gives
\beq
\label{eq:delta_squared_basso}
\Delta^2=J^2+S
\(
A_1\sqrt{\lambda}+A_2+\dots
\)
+S^2
\(
B_1+\frac{B_2}{\sqrt\lambda}
+\dots
\)
+S^3
\(
\frac{C_1}{\lambda^{1/2}}
+\frac{C_2}{\lambda}
+\dots
\)
+{\cal O}({ S}^4)\;,
\eeq
where the coefficients $A_i,\;B_i,\;C_i$ are some functions of $J$. There exists a one-to-one mapping between the coefficients $\alpha_i$, $\beta_i$, etc. and $A_i$, $B_i$ etc, which is rather complicated but easy to find. We note that this structure of $\Delta^2$ coincides with Basso's conjecture in \cite{Basso:2011rs} for mode number $n=1$ \footnote{The generalization of \eq{eq:delta_squared_basso} for $n>1$ is not fully clear, as noted in \cite{Gromov:2011bz}, and this case will be discussed in appendix \ref{sec:appN}.}. The pattern in \eq{eq:delta_squared_basso} continues to higher orders in $S$ with further coefficients $D_i$, $E_i$, etc. and powers of $\lambda$ suppressed incrementally. This structure is a nontrivial constraint on $\Delta$ itself as one easily finds from \eq{eq:delta_squared_basso} that
\beqa\la{eq:delta_basso}
\Delta&=&J+\frac{S}{2J}
\(
A_1\sqrt{\lambda}+A_2+\frac{A_3}{\sqrt{\lambda}}+\dots
\)\\
\nn&+&S^2
\(
- \frac{A_1^2}{8J^3} \, \lambda
-  \frac{A_1A_2}{4J^3} \, \sqrt{\lambda}
+\[\frac{B_1}{2J}-\frac{A_2^2+2A_1 A_3}{8J^3}\]
+
\[
\frac{B_2}{2J}
-\frac{A_2A_3+A_1A_4}{4J^3}
\]  \frac{1}{\sqrt\lambda}
+\dots
\).
\eeqa
By definition the coefficients of $S$ and $S^2$ are the slope and curvature functions respectively, so now we have their expansions at strong coupling in terms of $A_i,\;B_i,\;C_i$, etc. Since the $S$ coefficient only contains the constants $A_i$, we can find all of their values by simply expanding the slope function \eq{eq:resultLO} at strong coupling. We get
\beq
\label{eq:bassos_as}
A_1=2\;\;,\;\;
A_2=-1\;\;,\;\;
A_3=J^2-\frac{1}{4}\;\;,\;\;
A_4=J^2-\frac{1}{4}\dots\;.
\eeq
Note that in this series the power of $J$ increases by two at every other member, which is a direct consequence of omitting odd powers of $\Delta$ from \eq{eq:sofdelta}. We also expect the same pattern to hold for the coefficients $B_i$, $C_i$, etc.

The curvature function written in terms of $A_i$, $B_i$, etc. is given by
\beqa
	\label{eq:ss_abc}
	\gamma^{(2)}_{J}(g) &=& -\frac{2 \pi ^2 g^2 A_1^2 }{J^3} - \frac{\pi g A_1 A_2 }{J^3}-\frac{A_2^2+2 A_1 A_3-4 B_1 J^2}{8 J^3} - \frac{A_2 A_3+A_1 A_4-2 B_2 J^2}{16 \pi g J^3} \\
	&-&  \frac{A_3^2+2 A_2 A_4+2 A_1 A_5-4 B_3 J^2}{128 \pi^2 g^2 J^3} - \frac{A_3 A_4 + A_2 A_5 + A_1 A_6 - 2 B_4 J^2}{256 \pi^3 g^3 J^3} + \mc{O}\left(\frac{1}{g^4}\right). \nonumber
\eeqa
The remaining unknowns here (up to order $1/g^4$) are $B_1$, $B_2$, which we expect to be constant due to the power pattern noticed above and $B_3$, $B_4$, which we expect to have the form $a J^2 + b$ with $a$ and $b$ constant.
These unknowns are immediately fixed by comparing the general curvature expansion \eq{eq:ss_abc} to the two explicit cases that we know for $J=2$ and $J=3$. We find
\beq
\label{eq:b1b2}
B_1=3/2\;\;,\;\;B_2=-3\,\zeta_3+\frac{3}{8},
\eeq
and
\beqa
\label{eq:b3b4}
	B_{3} = -\frac{J^2}{2}-\frac{9 \, \zeta_3}{2}+\frac{5}{16} \;\;,\;\; B_{4} = \frac{3}{16} J^2 (16 \, \zeta_3+20 \, \zeta_5-9)-\frac{15 \, \zeta_5}{2}-\frac{93 \, \zeta_3}{8}-\frac{3}{16}.
\eeqa
Having fixed all the unknowns we can write the strong coupling expansion of the curvature function for arbitrary values of $J$ as
\beqa
 \gamma^{(2)}_{J}(g) &=& -\frac{8 \pi ^2 g^2}{J^3}+\frac{2 \pi  g}{J^3}+\frac{1}{4 J}+\frac{1-J^2 (24 \, \zeta_3 +1)}{64 \pi  g J^3} - \frac{8 J^4+J^2 (72 \, \zeta_3 +11)-4}{512 g^2 \left(\pi ^2 J^3\right)}  \nonumber \\
  &+& \frac{3 \left(8 J^4 (16 \, \zeta_3 +20 \, \zeta_5-7)-16 J^2 (31 \, \zeta_3 +20 \, \zeta_5+7)+25\right)}{16384 \pi ^3 g^3 J^3} + \mc{O}\left(\frac{1}{g^4}\right).
\eeqa
Expanding $\gamma^{(2)}_{J=4}$ defined in \eq{gamma2L4} at strong coupling numerically we were able to confirm the above result with high precision.

\subsection{Anomalous dimension of short operators}
\label{sec:Konishidimension}

In this section we will use the knowledge of slope functions $\gamma^{(n)}_J$ at strong coupling to find the strong coupling expansions of scaling dimensions of operators with finite $S$ and $J$, in particular we will find the three loop coefficient of the Konishi operator by utilizing the techniques of
\cite{Basso:2011rs,Gromov:2011bz}. What follows is a quick recap of the main ideas in these papers.

We are interested in the coefficients of the strong coupling expansion of $\Delta$, namely
\beq
	\Delta = \Delta^{(0)} \lambda^\frac{1}{4} + \Delta^{(1)} \lambda^{-\frac{1}{4}}  + \Delta^{(2)} \lambda^{-\frac{3}{4}} + \Delta^{(3)} \lambda^{-\frac{5}{4}} + \dots
\eeq
First, we use Basso's conjecture \eq{eq:delta_squared_basso} and by fixing $S$ and $J$ we re-expand the square root of $\Delta^2$ at strong coupling to find
\beq
	\label{eq:delta_abc}
	\Delta = \sqrt{A_1 S} \, \sqrt[4]{\lambda}  + \frac{\sqrt{A_1} \left( J^2 + A_2 S + B_1 S^2 \right)}{2 A_1 \sqrt{S}} \, \frac{1}{\sqrt[4]{\lambda}} + {\cal O}\(\frac{1}{\lambda^\frac{3}{4}}\).
\eeq
Thus we reformulate the problem entirely in terms of the coefficients $A_i$, $B_i$, $C_i$, etc. For example, the next coefficient in the series, namely the two-loop term is given by
\beq
	\label{eq:delta_2loops_abc}
	\Delta^{(2)} = -\frac{\left(2 A_2 + 4 B_1+J^2\right)^2-16 A_1 (A_3+2 B_2+4 C_1)}{16 \sqrt{2} A_2^{3/2}}.
\eeq
Further coefficients become more and more complicated, however a very clear pattern can be noticed after looking at these expressions: we see that the term $\Delta^{(n)}$ only contains coefficients with indices up to $n+1$, e.g. the tree level term $\Delta^{(0)}$  only depends on $A_1$, the one-loop term depends on $A_1$, $A_2$, $B_1$, etc. Thus we can associate the index of these coefficients with the loop level. Conversely, from the last section we learned that the letter of $A_i$, $B_i$, etc. can be associated with the order in $S$, i.e. the slope function fixed all $A_i$ coefficients and the curvature function in principle fixes all $B_i$ coefficients.

\subsubsection{Matching with classical and semiclassical results}

Looking at \eq{eq:delta_abc} we see that knowing $A_i$ and $B_i$ only takes us to one loop, in order to proceed we need to know some coefficients in the $C_i$ and $D_i$ series. This is where the next ingredient in this construction comes in, which is the knowledge of the classical energy and its semiclassical correction in the Frolov-Tseytlin limit, i.e. when $\mc S \equiv S/\sqrt\lambda$ and $\mc J \equiv J/\sqrt\lambda$ remain fixed, while $S$, $J$, $\lambda \rightarrow \infty$. Additionally we will also be taking the limit $\cal{S} \rightarrow \mathrm{0}$ in all of the expressions that follow. In particular, the square of the classical energy has a very nice form in these limits and is given by \cite{Gromov:2011de,Gromov:2011bz}
\beqa
 \label{delta_tree}
 {\cal D}_{\rm classical}^2&=&{\cal J}^2+2 \, {\cal S} \, \sqrt{{\cal J}^2+1}+{\cal S}^2 \, \frac{2 {\cal J}^2+3}{2
   {\cal J}^2+2}-{\cal S}^3 \, \frac{{\cal J}^2+3}{8
   \left({\cal J}^2+1\right)^{5/2}}
   +{\cal O}\left({\cal S}^4\right)\;,
\eeqa
where ${\cal D}_{\rm classical} \equiv \Delta_{classical} / \sqrt{\lambda}$. The 1-loop correction to the classical energy is given by
\beqa
\label{delta_oneloop}
\Delta_{sc} \simeq
\frac{-{\cal S}}{2 \left({\cal J}^3+{\cal J}\right)}+{\cal S}^2\[\frac{3 {\cal J}^4+11 {\cal J}^2+17
   }{16 {\cal J}^3 \left({\cal J}^2+1\right)^{5/2}}
\!-\!\sum_{\substack{m>0 \\ m\neq n}}\frac{n^3m^2  \left(2 m^2+n^2 {\cal J}^2-n^2\right)}{{\cal J}^3 \left(m^2-n^2\right)^2
   \left(m^2+n^2 {\cal J}^2\right)^{3/2}}\]
\eeqa
If the parameters $\cal{S}$ and $\cal{J}$ are fixed to some values then the sum can be evaluated explicitly in terms of zeta functions. We now add up the classical and the 1-loop contributions\footnote{Note that they mix various orders of the coupling.}, take $S$ and $J$ fixed at strong coupling and compare the result to \eq{eq:delta_squared_basso}. By requiring consistency we are able to extract the following coefficients,
$$
 \label{eq:abcd2}
 \begin{array}{rcrlrlrcl}
  A_1 &=&  &2, &A_2&  &=& -&1  \\
  B_1 &=&  &3/2, &B_2&  &=& -&3\,\zeta_3+\frac{3}{8}  \\
  C_1 &=& -&3/8, &C_2& &=& &\frac{3}{16} \, (20 \, \zeta_3 + 20 \, \zeta_5 - 3) \\
  D_1 &=&  &31/64, &D_2& &=& & \frac{1}{512} (-4720 \, \zeta_3 - 4160 \, \zeta_5 -2520 \, \zeta_7 +81)
 \end{array}
$$
As discussed in the previous section, we can in principle extract all coefficients with indices $1$ and $2$. In order to find e.g. $B_3$ we would need to extend the quantization of the classical solution to the next order. Note that the coefficients $A_1$, $A_2$ and $B_1$, $B_2$ have the same exact values that we extracted from the slope and curvature functions.

\subsubsection{Result for the anomalous dimensions at strong coupling}

\begin{table}[t]
\begin{tabular}{|l||rl|l|l||l|l|l|}
  \hline
  $(S,J)$ & \multicolumn{2}{|l|}{$\lambda^{-5/4}$ prediction} & $\lambda^{-5/4}$ fit & error & fit order\\
  \hline
  $(2,2)$ & $\frac{15 \, \zeta_5}{2} + 6 \, \zeta_3+\frac{1}{2}$&$= 15.48929958$ & $14.12099034$ & $9.69\%$ & 6\\
  $(2,3)$ & $\frac{15 \, \zeta_5}{2} + \frac{63 \, \zeta_3}{8} - \frac{619}{512}$&$= 16.03417190$ & 14.88260078 & $7.74\%$ & 5 \\
  $(2,4)$ & $\frac{21 \, \zeta_3}{2} + \frac{15 \, \zeta_5}{2} - \frac{17}{8}$&$= 18.27355565$ & $16.46106336$ & $11.0\%$ & 7\\
  \hline
\end{tabular}
\caption{Comparisons of strong coupling expansion coefficients for $\lambda^{-5/4}$ obtained from fits to TBA data versus our predictions for various operators. The fit order is the order of polynomials used for the rational fit function (see \cite{Gromov:2011bz} for details).}
\label{tab:coefficients}
\end{table}

The key observation in \cite{Gromov:2011bz} was that once written in terms of the coefficients $A_i$, $B_i$, $C_i$, the two-loop term $\Delta^{(2)}$ only depends on $A_{1,2,3}$, $B_{1,2}$, $C_{1}$ as can be seen in \eq{eq:delta_2loops_abc}. As discussed in the last section, the one-loop result fixes all of these constants except $A_3$, which in principle is a contribution from a true two-loop calculation. However we already fixed it from the slope function and thus we are able to find
\beq
	\Delta^{(2)} = \frac{-21\,S^4 +(24-96\,\zeta_3) S^3+4 \left(5 J^2-3\right) S^2+8 J^2 S -4 J^4}{64 \sqrt{2}\,S^{3/2}}.
\eeq
Now that we know the strong coupling expansion of the curvature function and thus all the coefficients $B_i$, we can do the same trick and find the three loop strong coupling scaling dimension coefficient $\Delta^{(3)}$, which now depends on $A_{1;2;3;4}$, $B_{1,2,3}$, $C_{1,2}$, $D_1$. We find it to be
\beqa
\label{D3anyS}
	\Delta^{(3)} &=& \frac{187\,S^6 + 6\,(208\,\zeta_3 + 160\,\zeta_5-43)\,S^5 +\left(-146\,J^2 - 4\,(336\,\zeta_3-41)\right)S^4 }{512 \sqrt{2}\,S^{5/2}} + \nonumber \\
	&+& \frac{\left(32\,(6\,\zeta_3+7)\,J^2-88\right)S^3 + \left(-28\,J^4 + 40\,J^2\right) S^2 - 24\,J^4 S + 8\,J^6}{512 \sqrt{2}\,S^{5/2}},
\eeqa
for $S=2$ it simplifies to
\beq
	\Delta^{(3)}_{S=2} = \frac{1}{512} \left(J^6-20 J^4+48 J^2 (4 \zeta_3 - 1)+64 (36 \, \zeta_3+60 \, \zeta_5+11)\right)
\eeq
and finally for the Konishi operator, which has $S=2$ and $J=2$ we get\footnote{The $\zeta_3$ and $\zeta_5$ terms are coming from semi-classics and were already known before \cite{Beccaria:2012xm} and match our result.}
 \beq
  \Delta^{(3)}_{S=2,J=2} = \frac{15 \, \zeta_5}{2} + 6 \, \zeta_3+\frac{1}{2}.
 \eeq
In order to compare our predictions with data available from TBA calculations \cite{Frolov:2010wt}, we employed Pad\'{e} type fits as explained in \cite{Gromov:2011bz}. The fit results are shown in table \ref{tab:coefficients}, we see that our predictions are within $\sim10\%$ error bounds, which is a rather good agreement. However we must be honest that for the $J=3$ and especially $J=4$ states we did not have as many data points as for the $J=2$ state and the fit is somewhat shaky.

\subsection{BFKL pomeron intercept}
\label{sec:bfkl}

The gauge theory operators that we consider in this paper are of high importance in high energy scattering amplitude calculations, especially in the Regge limit of high energy and fixed momentum transfer
\cite{ReggeOriginal,Gribov}. In this limit one can approximate the scattering amplitude as an exchange of effective particles, the so-called reggeized gluons, compound states of which are frequently called \emph{pomerons}. When momentum transfer is large, perturbative computations are possible and the so-called `hard pomeron' appears, the BFKL pomeron \cite{bfkl}. The BFKL pomeron leads to a power law behaviour of scattering amplitudes $s^{j(\Delta)}$, where $j(\Delta)$ is called the Reggeon spin and $s$ is the energy transfer of the process. The remarkable connection between the pomeron and the operators we consider can be symbolically stated as
\beq
	\mathrm{pomeron} = \Tr\(Z \; \cD_+^S \; Z \)+\dots
\eeq
where we are now considering twist two operators ($J=2$) and the spin $S$ can take on complex values by analytic continuation. The Reggeon spin $j(\Delta)$ (also refered to as a Regge trajectory) is a function of the anomalous dimension of the operator and is related to spin $S$ as $j(\Delta) = S(\Delta) + 2$. Some of these trajectories are shown in figure \ref{pic:bfkl}. A very important quantity in this story is the BFKL intercept $j(0)$, which we consider next.

\FIGURE[ht]{
\label{pic:bfkl}
    \begin{tabular}{cc}
    \includegraphics[scale=0.7]{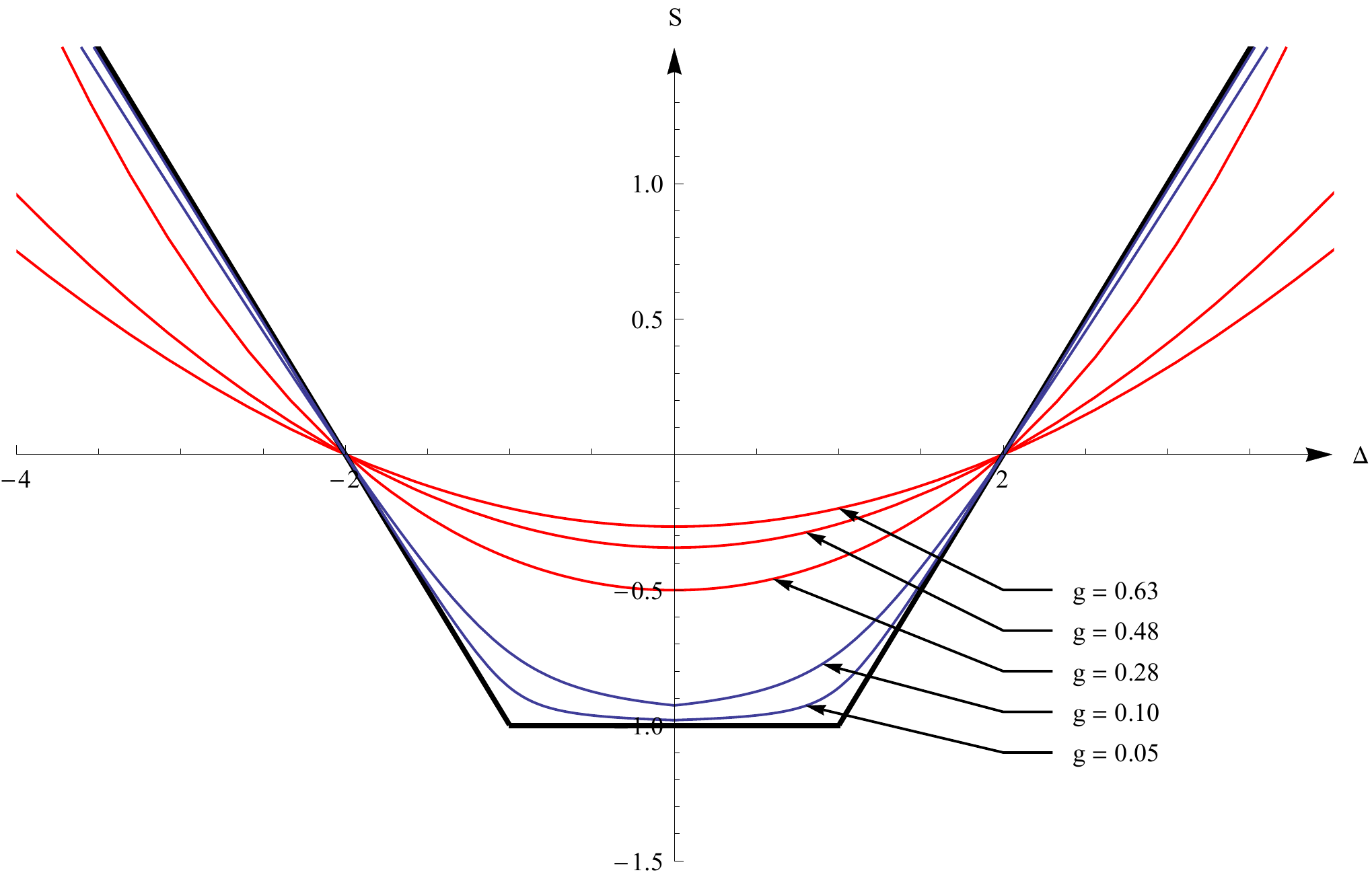}\\
    \end{tabular}
\caption{The BFKL trajectories $S(\Delta)$ at various values of the coupling. Blue lines are obtained using the known two loop weak coupling expansion \cite{Brower:2006ea,Kotikov:2002ab} and red lines are obtained using the strong coupling expansion \cite{Costa:2012cb,Kotikov:2013xu,Brower:2013jga}.}
}

\FIGURE[ht]{
\label{pic:intercept}
    \begin{tabular}{cc}
    \includegraphics[scale=0.7]{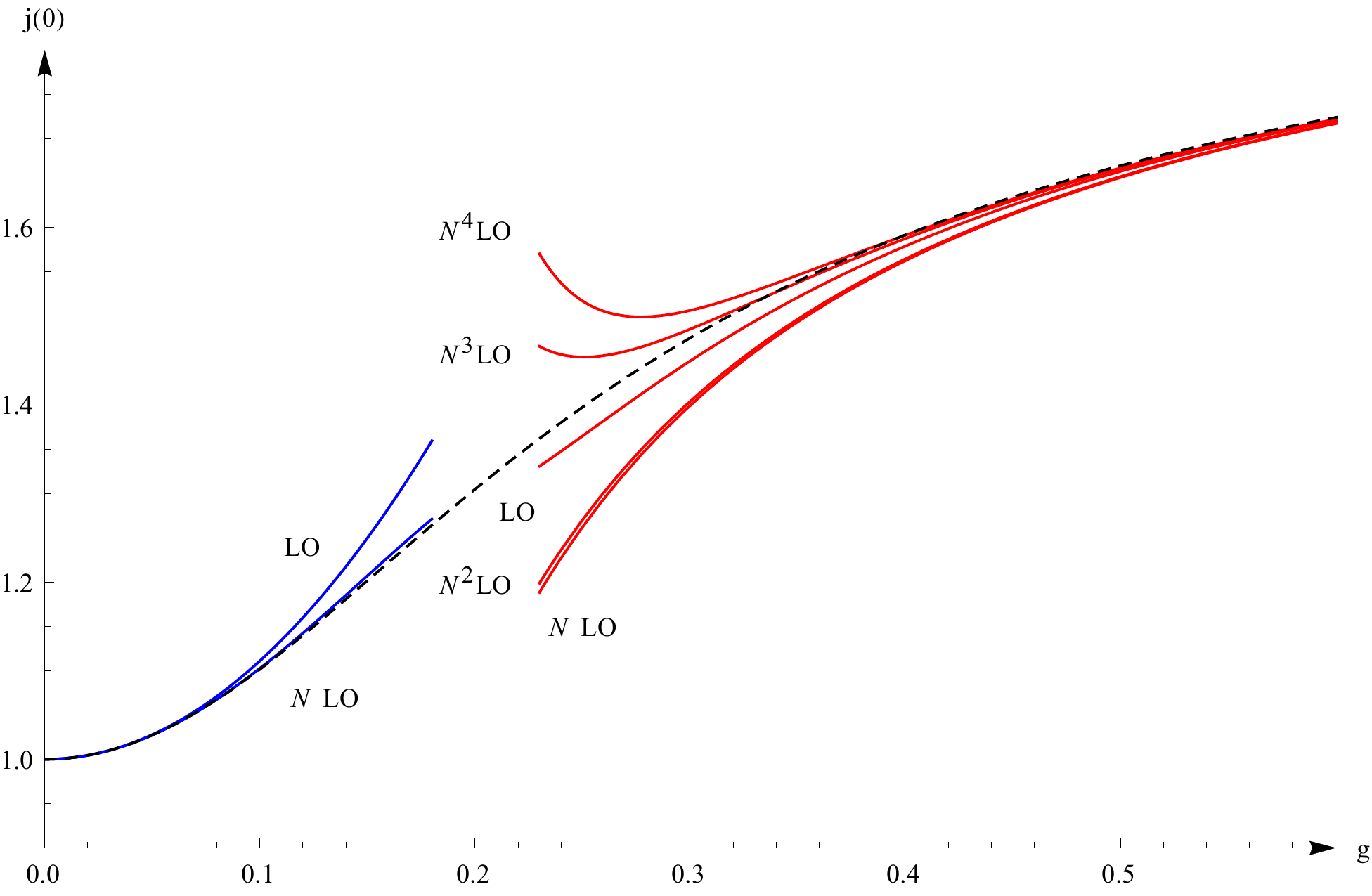}\\
    \end{tabular}
\caption{The BFKL intercept $j(0) = 2 + S(0)$ dependence on the coupling constant $g$ at two orders at weak coupling (blue lines), four orders at strong coupling (red lines) and a Pad\'{e} type interpolating function in between (dashed line).}
}

One can also use the same techniques as in the previous section to calculate the strong coupling expansion of the BFKL intercept. As stated before, the intercept of a BFKL trajectory $j(\Delta)$ is simply $j(0)$ and we already wrote down an ansatz for $S(\Delta)$ in \eq{eq:sofdelta}. The coefficients $\alpha_i$, $\beta_i$, etc. are in one-to-one correspondence with the coefficients $A_i$, $B_i$ etc. from \eq{eq:delta_basso}, values of which we found in the previous sections. Plugging in their values we find

\beq
	\alpha_1=1/2,\ \alpha_2=1/4,\ \alpha_3=-1/16\ ,\
	\alpha_4=-\frac{3\zeta_3}{2}-\frac{1}{2},\
	\eeq
\beq
	\alpha_5=-\frac{9 \zeta_3}{2}-\frac{361}{256},\
	\alpha_6=-\frac{39 \zeta_3}{4}-\frac{511}{128}
\eeq
\beq
	\beta_3=-3/16,\ \beta_4=\frac{3\zeta_3}{8}-\frac{21}{64},\
	\beta_5=\frac{9 \zeta_3}{8}-\frac{51}{128},\
	\beta_6=\frac{45 \zeta_3}{8}+\frac{15 \zeta_5}{16}+\frac{141}{512}
\eeq
\beq
	\gamma_5=\frac{21}{128},\ \gamma_6=-\frac{51 \zeta_3}{64}-\frac{15 \zeta_5}{64}+\frac{129}{256}
\eeq
Furthermore, setting $\Delta = 0$ we find the intercept to be
\beqa \nn
j(0) = 2 + S(0) &=& 2 -\frac{2}{\lambda^{1/2}}-\frac{1}{\lambda }+ \frac{1}{4\,\lambda^{3/2}}+\left(6 \zeta_3+2\right) \frac{1}{\lambda^2} \\
	&+& \left(18 \, \zeta_3 + \frac{361}{64} \right) \frac{1}{\lambda^{5/2}} + \left(39 \, \zeta_3 + \frac{511}{32}\right) \frac{1}{\lambda^3}  + \mathcal{O}\left(\frac{1}{\lambda^{7/2}}\right).
\eeqa
The first four terms successfully reproduce known results \cite{Costa:2012cb,Kotikov:2013xu,Brower:2013jga} and the last two terms of the series are a new prediction (their derivation relies on the knowledge of the constants $B_{3,4;J=2}$ found in the last section). On Figure \ref{pic:intercept} we show
plots of the intercept at weak and at strong coupling.

\section{Conclusions}

In this paper we applied the recently proposed $\bP\mu$-system of Riemann-Hilbert type equations to study anomalous dimensions in the $sl(2)$ sector of planar $\cN=4$ SYM theory. Our main result are the expressions \eq{gamma2L2}, \eq{gamma2L3} and \eq{gamma2L4} for the curvature function $\gamma_J^{(2)}(g)$, i.e. the coefficient of the $S^2$ term in the anomalous dimension at small spin $S$. These results correspond to operators with twist $J=2,3$ and $4$. Curiosly, we found that they involve essential parts of the BES dressing phase in the integral representation.

We derived these results by solving the $\bP\mu$-system to order $S^2$ and they are exact at any coupling.  While expansion in small $S$ (but at any coupling)
seems hardly possible to perform in the TBA approach, here it resembles a perturbative expansion -- the $\bP\mu$-system is solved order by order in $S$ and the coupling is kept arbitrary.

For $J=2$ and $J=3$ our calculation perfectly matches known results to four loops at weak coupling. This includes in particular the leading finite-size correction at $J=2$. At strong coupling we obtained the expansion of our results numerically, and also found full agreement with known predictions. This provides yet another check that our result incorporates all wrapping corrections. Going to higher orders in this expansion we were able to use the EZFace calculator \cite{ezface} to fit
 the coefficients as linear combinations of $\zeta_3$ and $\zeta_5$ (and confirmed the outcome with high precision). By combining these coefficients with the other known results, we obtained the 3-loop coefficient in the Konishi anomalous dimension at strong coupling. This serves as a highly nontrivial prediction for a direct string theory calculation, which hopefully may be done along the lines of \cite{Vallilo:2011fj, Roiban:2011fe}. Our results also predict the value of two new coefficients for the pomeron intercept at strong coupling.

For the future analysis it would be interesting to build an integral equation which would generate iteratively small $S$ corrections
and to be able to approach finite values of $S$ at any coupling. Furthermore, extension of this approach to the boundary problems,
twisted boundary conditions and even q-deformations \cite{Bajnok:2013wsa, Bajnok:2012xc, betadef, Arutyunov:2012ai, Arutyunov:2012zt} would give a rich set of new analytical results.
Finally, applications of our methods to ABJM model \cite{Lambert,ABJM}
and comparison with the localization results \cite{Kapustin:2009kz, Drukker:2010nc, Drukker:2011zy} would give the unknown interpolation function
$h(\lambda)$, the only ingredient missing in the integrability framework \cite{GVABA,ABJMTBA1,ABJMTBA2}.
As the $\bP\mu$-system \cite{pmuABJM} for ABJM model has various peculiar features compared to ${\cal N}=4$ SYM
 it would be especially interesting to study this case.

\section*{Acknowledgements}
We are grateful to M.~Alfimov, A.~Gorsky, V.~Kazakov, A.~Kotikov,
J.~Penedones,
A.~Tseytlin and D.~Volin for useful discussions. The research leading to these results has received funding from the People Programme (Marie Curie Actions) of the European Union's Seventh Framework Programme FP7/2007-2013/ under REA Grant Agreement No 317089 (GATIS).
N.G. is grateful to the Holograv Conference ``Gauge/Gravity Duality", MPI for Physics, Munich, 15-19 July 2013 where a part of this work was done for hospitality.
The work of F.~L.-M. is supported in part by the grants RFBR-12-02-00351-a and PICS-12-02-91052. We also wish to thank the STFC for partial support from the consolidated grant ST/J002798/1.

\appendix

\section{Summary of notation and definitions}
In this appendix we summarize some notation used throughout the paper.
\paragraph{Laurent expansions in $x$}
$\\$
$\\$
We often represent functions of the spectral parameter $u$ as a series in $x$
\beq
f(u)=\sum\limits_{n=-\infty}^{\infty}f_n x^n
\eeq
with
\beq
	u=g(x+1/x).
\eeq
We denote by $[f]_+$ and $[f]_-$ part of the series with positive and negative powers of $x$:
\beqa
&&[f]_+=\sum\limits_{n=1}^{\infty}f_n x^n, \\
&&[f]_-=\sum\limits_{n=1}^{\infty}f_{-n} x^{-n}.
\eeqa

As a function of $u$, $x(u)$ has a cut from $-2g$ to $2g$. For any function $f(u)$ with such a cut we denote another branch of $f(u)$ obtained by analytic continuation (from $\Im \;u>0$) around the branch point $u=2g$ by $\tilde f(u)$. In particular, $\tilde x=1/x$.

\paragraph{Functions $\sinh_\pm$ and $\cosh_\pm$}
\par

$\\$

We define $I_k=I_k(4 \pi g)$, where $I_k(u)$ is the modified Bessel function of the first kind.
Then
 \beqa
&& \sinh_+=[\sinh(2\pi u)]_+=\sum\limits_{k=1}^\infty I_{2k-1}x^{2k-1}, \\
&& \sinh_-=[\sinh(2\pi  u)]_-=\sum\limits_{k=1}^\infty I_{2k-1}x^{-2k+1},\\
&& \cosh_+=[\cosh(2\pi u)]_+=\sum\limits_{k=1}^\infty I_{2k}x^{2k}, \\
&& \cosh_-=[\cosh(2\pi u)]_-=\sum\limits_{k=1}^\infty I_{2k}x^{-2k}.
 \eeqa
 In some cases we denote for brevity
\beq
	\sh_-^x=\sinh_-(x),\ \ \ \ch_-^x=\cosh_-(x).
\eeq

\paragraph{Integral kernels}
$\\$
$\\$
In order to solve for $\bP_a^{(1)}$ in section \ref{sec:CalculationofPa} we introduce integral operators $H$ and $K$ with kernels
\beqa
H(u,v)&=&-\frac{1}{4\pi i}\frac{\sqrt{u-2g}\sqrt{u+2g}}{\sqrt{v-2g}\sqrt{v+2g}}\frac{1}{u-v}dv, \\
K(u,v)&=&+\frac{1}{4\pi i}\frac{1}{u-v}dv,
\eeqa
which satisfy
\beq
\tilde f+f=h\;\;,\;\;f=H\cdot h\;\;\;\;{\rm and}\;\;\;\;
\tilde f-f=h\;\;,\;\;f=K\cdot h.
\label{lab123}
\eeq

Since the purpose of $H$ and $K$ is to solve equations of the type \ref{lab123}, $H$ usually acts on functions $h$ such that $\tilde h=h$, whereas $K$ acts on $h$ such that $\tilde h=-h$. On the corresponding  classes of functions, provided also that the constant term in their Laurent expansion (denoted as $[h]_0$) is zero, $H$ and $K$ can be represented by kernels which are equal up to a signЖ
\beqa
H(u,v)&=&-\left.\frac{1}{2\pi i}\frac{1}{x_u-x_v}dx_v\right|_{\tilde h=h},\;\;\\
K(u,v)&=&\left.\frac{1}{2\pi i}\frac{1}{x_u-x_v}dx_v\right|_{\tilde h=-h}.
\eeqa
In order to be able to deal with series in half-integer powers of $x$ in section \ref{sec:SolvingPmuL3} we introduce modified kernels:
\beqa
&&H^*\cdot f\equiv\frac{x+1}{\sqrt{x}}H\cdot\frac{\sqrt{x}}{x+1} f, \\
&&K^*\cdot f\equiv\frac{x+1}{\sqrt{x}}K\cdot\frac{\sqrt{x}}{x+1} f.
\eeqa
Finally, to write the solution to equations of the type \eqref{eq:mudiscNLO}, we introduce the operator $\Gamma'$ and its more symmetric version $\Gamma$

\beq
\(\Gamma'\cdot h\)(u)\equiv \oint_{-2g}^{2g}\frac{dv}{{4\pi i}}\d_u \log \frac{\Gamma[i (u-v)+1]}{\Gamma[-i (u-v)]}h(v),
\eeq

\beq
\(\Gamma\cdot h\)(u)\equiv \oint_{-2g}^{2g}\frac{dv}{{4\pi i}}\d_u \log \frac{\Gamma[i (u-v)+1]}{\Gamma[-i (u-v)+1]}h(v).
\eeq
\paragraph{Periodized Chebyshev polynomials}
\label{sec:appPeriodized}
$\\$
$\\$
Periodized Chebyshev polynomials appearing in $\mu_{ab}^{(1)}$ are defined as
\beqa
&&p_a'(u)=\frac{1}{2}\Sigma\cdot\left[x^a+1/x^a\right]=\Sigma\cdot\left[T_a\(\frac{u}{2g}\)\right],\\
&&p_a(u)=p_a'(u)+\frac{1}{2}\(x^a(u)+x^{-a}(u)\),
\eeqa
where $T_a(u)$ are Chebyshev polynomials of the first kind. Here is the explicit form for the first five of them:	
\beqa
&&p_0'=-i(u-i/2),\\
&&p_1'=-i\frac{u(u-i)}{4g},\\
&&p_2'=-i\frac{(u-i/2)(-6g^2+u^2-iu)}{6g^2},\\
&&p_3'=-i\frac{ u (u-i) \left(-6 g^2+u (u-i)\right)}{8 g^3},\\
&&p_4'=-i\frac{ \left(u-\frac{i}{2}\right) \left(30 g^4-20 g^2 u^2+20 i g^2 u+3 u^4-6 i u^3-2 u^2-i u\right)}{30 g^4}.
\eeqa

\section{The slope function for odd $J$}
\label{sec:oddL}

Here we give details on solving the $\bP\mu$-system for odd $J$ at leading order in the spin. First, the parity of the $\mu_{ab}$ functions is different from the even $J$ case, which can be seen from the asymptotics \eq{eq:muasymptotics}. Following arguments similar to the discussion for even $J$ in section \ref{sec:evenLsol}, we obtain
\beq
	\mu_{12}=1,\ \mu_{13}=0,\ \mu_{14}=0,\  \mu_{24}=\cosh(2\pi u),\ \mu_{34}=1.
\eeq
Plugging these $\mu_{ab}$ into \eqref{eq:pmuexpanded} we get a system of equations for $\bP_a$
\beqa
&&\tilde \bP_1= -\bP_3,  \\
&&\tilde \bP_2= -\bP_4 -\bP_1 \cosh(2\pi u), \\
&&\tilde \bP_3= -\bP_1,\\
&&\tilde \bP_4= -\bP_2+\bP_3 \cosh(2\pi u).
\eeqa
This system can be solved in a similar way to the even $J$ case. The only important difference is that due to asymptotics \eq{eq:asymptotics} the $\bP_a$ acquire an extra branch point at $u=\infty$.

Let us first rewrite the equations for $\bP_1,\bP_3$ as
\beqa
\tilde\bP_1+\tilde\bP_3&=&-\(\bP_1+\bP_3\)\\
\tilde\bP_1-\tilde\bP_3&=&\bP_1-\bP_3.
\eeqa
This, together with the asymptotics \eqref{eq:asymptotics} implies $\bP_1=\epsilon  x^{-J/2},\;\bP_3=-\epsilon  x^{J/2}$ where $\epsilon$ is a constant. Let us note that these $\bP_1, \bP_3$ contain half-integer powers of $x$, and the analytic continuation around the branch points at $\pm 2g$ replaces $\sqrt{x}\to1/\sqrt{x}$. Now, taking the sum and difference of the equations for $\bP_2,\;\bP_4$ we get
\beqa
&&\tilde\bP_2+\tilde\bP_4+\bP_2+\bP_4=-a_1\(x^{J/2}+x^{-J/2}\)\cosh{2\pi u}\\
&&\tilde\bP_2-\tilde\bP_4-\(\bP_2-\bP_4\)=a_1\(x^{J/2}-x^{-J/2}\)\cosh{2\pi u}
\eeqa
We can split the expansion
\beq
	\cosh{2\pi u}=\sum\limits_{k=-\infty}^{\infty}I_{2k}x^{2k}
\eeq	
into the positive and negative parts according to
\beq
\cosh{2\pi u}=\cosh_-+\cosh_++I_0
\eeq
where
\beq
\cosh_+=\sum\limits_{k=1}^{\infty}I_{2k}x^{2k},\;\ \ \ \ \cosh_-=\sum\limits_{k=1}^{\infty}I_{2k}x^{-2k}.
\eeq
Then we can write
\beqa
&&\bP_2+\bP_4=-a_1(x^{J/2}+x^{-J/2})\cosh_--a_1 I_0 x^{-J/2}+Q, \\
&&\bP_2-\bP_4=-a_1(x^{J/2}-x^{-J/2})\cosh_-+a_1 I_0 x^{-J/2}+P,
\eeqa
where $Q$ and $P$ are some polynomials in $\sqrt{x},1/\sqrt{x}$ satisfying
\beq\label{QP}
	\tilde Q=-Q,\; \tilde P=P.
\eeq
We get
\beqa
\label{eq:P2tmp}
&&\bP_2=-a_1 x^{J/2}\cosh_- +\frac{Q+P}{2},\\
\label{eq:P4tmp}
&&\bP_4=a_1 x^{-J/2}\cosh_- -a_1 I_0 x^{-J/2}+\frac{Q-P}{2}.
\eeqa
Now imposing the correct asymptotics of $\bP_2$ we find
\beq
\frac{P+Q}{2}=a_1 x^{J/2}\sum\limits_{k=1}^{\frac{J-1}{2}}I_{2k}x^{-2k}
\eeq
Due to \eq{QP} this relation fixes $Q$ and $P$ completely,
and we obtain the solution given in section 2.1,
\beqa
\label{eq:musolLOoddL}
&&	\mu_{12}=1,\ \mu_{13}=0,\ \mu_{14}=0, \ \mu_{24}=\cosh(2\pi u),\ \mu_{34}=1, \\
&&   \bP_1=a_1 x^{-J/2}, \\
&&   \bP_2=-a_1 x^{J/2}\sum\limits_{k=-\infty}^{-\frac{J+1}{2}}I_{2k}x^{2k},\\
&&   \bP_3=-a_1 x^{J/2}, \\
\label{eq:P4solLOoddL}
&&    \bP_4=a_1 x^{-J/2}\cosh_--a_1 x^{-J/2}\sum\limits_{k=1}^{\frac{J-1}{2}}I_{2k}x^{2k}-a_1 I_0 x^{-J/2}.
\eeqa
Notice that the branch point at infinity is absent from the product of any two $\bP$'s, as it should be \cite{PmuPRL}, \cite{PmuLong}. One can check that this solution gives again the correct result \eqref{eq:resultLO} for the slope function.

\section{NLO solution of $P\mu$ system: details}
\label{sec:NLOapp}
In this appendix we will provide more details on the solution of the $\bP\mu$-system and calculation of curvature function for $J=2,3,4$ which was presented in the main text in sections \ref{sec:SolvingPmuL2}, \ref{sec:SolvingPmuL3}, \ref{sec:SolvingPmuL4}.

\subsection{NLO corrections to $\mu_{ab}$ for $J=2$}
\label{sec:appmu2}
Here we present some details of calculation of NLO corrections to $\mu_{ab}$ for $J=2$ omitted in the main text. As described in section \ref{sec:muNLOL2}, $\mu^{(1)}_{ab}$ are found as solutions of \eqref{eq:mudiscNLO} with appropriate asymptotics. The general solution of this equation consists of a general solution of the corresponding homogeneous equation (which can be reduced to one-parametric form \eqref{eq:periodicpart}) and a particular solution of the inhomogeneous one. The latter can be taken to be
\beq
\mu_{ab}^{disc}=\Sigma\cdot\(\bP_a^{(1)} \tilde\bP_b^{(1)}- \bP_b^{(1)} \tilde\bP_a^{(1)}\).
\eeq
One can get rid of the operation $\Sigma$, expressing $\mu_{ab}^{disc}$ in terms of $\Gamma'$ and $p_a'$. This procedure is based on two facts: the definition \eqref{paprime} of $p'_a$ and the statement that on functions decaying at infinity $\Sigma$ coincides with $\Gamma'$ defined by \eqref{Gammaprime}. After a straightforward but long calculation we find
\beqa
&&\mu_{31}^{disc}=\epsilon^2\Sigma\(\frac{1}{x^2}-x^2\)=
-\epsilon^2\;\;\(\Gamma\cdot x^2+p_2\),\\
&&\mu_{41}^{disc}=
   \epsilon^2\[-2I_1p_1-4I_1\Gamma\cdot x+\sinh(2\pi u)\(\Gamma\cdot x^2+p_0\)+
   \Gamma\cdot\sinh_-\(x-\frac{1}{x}\)^2\],\\
&&\mu_{43}^{disc}=
   -2{\epsilon^2}\[-2I_1p_1-4 I_1\Gamma\cdot x+\sinh(2\pi u)(p_2-p_0)+\Gamma\cdot\sinh_-\(x-\frac{1}{x}\)^2
   \],\\
&&\mu_{21}^{disc}=
   \epsilon^2\[2 I_1\Gamma\cdot x-\sinh(2\pi u)\;\Gamma\cdot x^2-\Gamma\cdot\sinh_-\(x^2+\frac{1}{x^2}\)
   \],\\
&&\mu_{24}^{disc}=
   \epsilon^2\[2I_1 \Gamma\cdot \sinh_- \(x+\frac{1}{x}\)+I_1^2p_0+ \right.\\
 && \left. +\sinh(2\pi u)\Gamma\cdot\sinh_-\(x^2-\frac{1}{x^2}\)-\Gamma\cdot \sinh_-^2\(x^2-\frac{1}{x^2}\)
   \].
\eeqa
Here we write $\Gamma$ and $p_a$ instead of $\Gamma'$ and $p'_a$ taking into account the discussion between equations \eqref{eq:ra} - \eqref{pa}.

\subsection{NLO solution of the $\bP\mu$-system at $J=3$}
\label{sec:appnlo3}
In this appendix we present some intermediate formulas for the calculation of curvature function for $J=3$ in section \ref{sec:SolvingPmuL3} omitted in the main text.
\begin{itemize}
\item
The particular solution of the inhomogeneous equation \eqref{eq:mudiscNLO} which we construct as $\mu_{31}^{disc}=\Sigma\cdot\(\bP_{a}^{(1)}\tilde\bP_{b}^{(1)}-\bP_{b}^{(1)}\tilde\bP_{a}^{(1)}\)$ can be written using the operation $\Gamma$  and $p_a$ defined by \eqref{pa} and \eqref{Gamma}\footnote{Alternatively one can use $p_a'$ and $\Gamma'$ instead of $p_a$ and $\Gamma$- see the discussion between the equations \eqref{eq:ra} - \eqref{pa}}
\beqa
&&\mu_{31}^{disc}=\Sigma\cdot({\bf P}_3 \tilde{\bf P}_1-{\bf P}_1 \tilde{\bf P}_3)=-2\epsilon^2\[\Gamma x^3+p_3\],\\
&&\mu_{41}^{disc}=
   -\epsilon^2\[2p_2I_2+2I_2\Gamma x^2+2\Gamma\cdot\cosh_-+(I_0-\cosh(2\pi u))p_0\],\\
&&\mu_{34}^{disc}=
   {\epsilon^2}\[2I_2\Gamma x +I_0\Gamma x^3-\Gamma\cdot(x^3+x^{-3})\cosh_-+\cosh(2\pi u)(2p_3+\Gamma x^3)\],\\
&&\mu_{21}^{disc}=
   \epsilon^2\[2I_2\Gamma x+(I_0-\cosh(2\pi u))\Gamma x^3-\Gamma\((x^3+x^{-3})\cosh(2\pi u)\)\],\\
&&\mu_{24}^{disc}=
   -2\epsilon^2\[-\frac{1}{2}\Gamma\cdot\cosh_-^2\(x^3-x^{-3}\)+\(\frac{\cosh(2\pi u)}{2}-I_0\)\Gamma\cdot\frac{\cosh_-}{x^3}\right.\\
   &&-I_2\Gamma\cdot\(x+\frac{1}{x}\)\cosh_--\frac{1}{2}\cosh(2\pi u)\Gamma\cdot x^3\cosh_-+\\&&\left.+\frac{I_0}{2}\(I_0-\cosh(2\pi u)\)\Gamma\cdot x^3+\frac{I_1 I_2}{2\pi g}\Gamma x-I_2^2 p_1 \].
\eeqa

\item
The zero mode of the system \eqref{P1L3}-\eqref{P4L3}, which we added to the solution in Eqs. \eq{P1J3}-\eq{P4J3} to ensure correct asymptotics, is
\beqa
\label{P1J3zm}
\bP_1^{\text{zm}}&=&L_1 x^{-1/2}+L_3x^{1/2},\\
\nn \bP_2^{\text{zm}}&=&-L_1 x^{1/2}\ch_-+L_2 x^{-1/2}-L_3x^{-1/2}\(\ch_-+\ofrac{2}I_0\)+L_4\(x^{1/2}-x^{-1/2}\),\\ \nn
\bP_3^{\text{zm}}&=&-L_1 x^{1/2}-L_3x^{1/2},\\ \nn
\bP_4^{\text{zm}}&=&-L_1\(I_0 x^{-1/2}+x^{-1/2}\cosh_-\)-L_2x^{1/2}+L_4(x^{1/2}-x^{-1/2})
\\ \nn
&-&L_3x^{1/2}\(\ch_-+\ofrac{2}I_0\).
\label{P4J3zm}
\eeqa
\end{itemize}

\subsection{NLO solution of the $\bP\mu$-system at $J=4$}
\label{sec:appL4}
Solution of the $\bP\mu$ system at NLO for $J=4$ is completely analogous to the case of $J=2$. The starting point is the LO solution \eqref{eq:P1solLOevenL}-\eqref{eq:P4solLOevenL}. As described in section \ref{sec:appmu2}, from LO $\bP_a$ we can find $\mu_{ab}$ at NLO. Its discontinuous part is
\label{sec:appnlo4}
\beqa
&&\mu_{31}^{disc}=
-\epsilon^2\;\;\(\Gamma\cdot x^4+p_4\),\\
&&\mu_{41}^{disc}=
\frac{1}{2} \epsilon ^2 \left(\sinh (2 \pi   u) \left(p_0+\Gamma \cdot x^4\right)+2 \left(I_1 p_1+I_3 p_3\right)+
\right.\\ 
\nn&&\left.
+\Gamma \cdot
  \sinh_- \left(x^2-\frac{1}{x^2}\right)^2-2 \left(I_1+I_3\right) \left(\Gamma \cdot x^3+\Gamma \cdot x\right)\right),\\
&&\mu_{43}^{disc}=\epsilon ^2 \left(\left(p_4-p_0\right) \sinh (2 \pi  u)+2 \left(I_1 p_1+I_3 p_3\right)-   \right.
\\ \nn
 &&\left.-\Gamma \cdot\sinh_-
   \left(x^2-\frac{1}{x^2}\right)^2
   +2 \left(I_1+I_3\right) \left(\Gamma \cdot x^3+\Gamma \cdot x\right)\right.\Bigg)
   ,\\ 
&&\mu_{21}^{disc}=\epsilon ^2 \left(-\frac{1}{2} \sinh (2 \pi  u) \Gamma \cdot x^4+I_1 p_3+I_3 p_1-
\right.
\\ \nn &&\left.
-\frac{1}{2} \Gamma \cdot \sinh_-
   \left(x^4+\frac{1}{x^4}\right)+I_1 \Gamma \cdot x^3+I_3 \Gamma \cdot x\right)
   ,\\
&&\mu_{24}^{disc}=\epsilon ^2 \left(\frac{1}{2} \sinh (2 \pi u) \Gamma \cdot\sinh_- \left(x^4-\frac{1}{x^4}\right)+I_3^2 p_2+I_1 I_3
   p_0-
      \right.
	\\ \nn&&\left.
   -\frac{1}{2} \Gamma \cdot \sinh_-^2 \left(x^4-\frac{1}{x^4}\right)
   +I_1 \Gamma \cdot \sinh_-
   \left(x^3+\frac{1}{x^3}\right)+   \right.
\\ \nn &&\left.+I_3 \Gamma \cdot\sinh_- \left(x+\frac{1}{x}\right)+\left(I_3^2-I_1^2\right)
   \Gamma \cdot x^2\right)
   ,
\eeqa
and as discussed for $J=2$ the zero mode can be brought to the form
\beqa
&& \pi_{12}=0,\ \pi_{13}=0,\ \pi_{14}=0,\\
 && \pi_{24}=c_{1,24}\cosh{2\pi u},\ \pi_{34}=0.
 \label{eq:periodicpartL4}
 \eeqa
After that, we calculate $r_a$ by formula \eqref{eq:ra} and solve the expanded to NLO $\bP\mu$ system for $\bP_a^{(1)}$ as
\beqa
&&\bP^{(1)}_3=H \cdot r_3, \\
&&\bP^{(1)}_1=\frac{1}{2}\bP^{(1)}_3+K\cdot \(r_1-\frac{1}{2}r_3\),\\
&&\bP^{(1)}_4=K\cdot\[(H\cdot r_3)\sinh(2\pi u)+r_4-\frac{1}{2}r_3\sinh(2\pi u) \]-C(x+1/x),\\
&&\bP^{(1)}_2=H\cdot\[-\bP^{(1)}_4-\bP^{(1)}_1\sinh(2\pi u)+r_2\]+C/x,
\eeqa
where $C$ is a constant which is fixed by requiring correct asymptotics of $\bP_2$.
Finally we find leading coefficients $A_a$ of $\P^{(1)}_a$ and use expanded up to ${\cal O}(S^2)$ formulas \eqref{AA1}, \eqref{AA2} in the same way as in section \ref{sec:resultL2} to obtain the result \eqref{gamma2L4}.

 \subsection{Result for $J=4$}
 \label{sec:SolvingPmuL4}
The final result for the curvature function at $J=4$ reads
\beqa
\label{gamma2L4}
&&\gamma^{(2)}_{J=4}=\oint \frac{du_x}{2\pi i}\oint \frac{du_y}{2\pi i}
\frac{1}{i g^2(I_3-I_5)^3} \left.\Bigg[ \right.\\ && \left.\nn
\frac{2 \(\sh_-^x\)^2 y^4 \left(I_3 \left(x^{10}+1\right)-I_5 x^2 \left(x^6+1\right)\right)}{x^4 \left(x^2-1\right)}-\frac{2 \(\sh_-^y\)^2 x^4 \left(y^8-1\right) \left(I_3 x^2-I_5\right)}{\left(x^2-1\right) y^4}+ \right.\\
&&+\frac{4 \sh_-^x \sh_-^y \left(x^4 y^4-1\right) \left(I_3+I_3 x^6 y^4-I_5 x^2 \left(x^2 y^4+1\right)\right)}{x^4 \left(x^2-1\right) y^4}\nn\\
&&+\left.\sh_-^y\(
\left(y^4+y^{-4}\right) x^{-1}\left(\left(I_1 I_5-I_3^2\right) \left(3 x^4+1\right)-2 I_1 I_3 x^6\right)+
\right.\right.\nn\\&&\left.\left.
+\frac{2 I_3 x^2
   \left(I_5 \left(x^2+1\right) x^2+I_1 \left(1-x^2\right)\right)-I_1 I_5 \left(x^2-1\right)^2+I_3^2 \left(-2 x^6+x^4+1\right)}{x(x^2-1)}+
   \right.\right.\nn\\&&\left.\left.
   +2
   \left(y^3+y^{-3}\right) \frac{I_1 I_3 x^6-I_1 I_5 x^4-I_3^2 \left(x^2-1\right)}{x^2-1}-
   \right.\right.\nn\\&&\left.\left.
   -2 I_3 \left(y+y^{-1}\right)
   \frac{ I_1\left(x^2-1\right)-I_3 \left(x^6-x^2+1\right)+I_5 \left(x^4-x^2+1\right)}{x^2-1}
   \)+
       \right.\nn\\&&\left.
   +\frac{4 x^6 y^2 I_3 \left(I_3^2-I_1^2\right)}{x^2-1}+\frac{4 x y I_1 \left(I_3 y^2+I_1\right) \left(I_3+I_5\right)}{x^2-1}+
   \right.\nn\\&&\left.
  \frac{2 y^4 \left(I_1+I_3\right) \left(I_1 I_5-I_3^2\right)}{x^2-1}
-\frac{2 y \left(y^2+1\right) \left(I_1+I_3\right) \left(I_1 I_5-I_3^2\right)}{x \left(x^2-1\right)}-
    \right.\nn\\&&\left.
-\frac{2 x^3 y \left(I_1+I_3\right) \left(I_1 \left(2 I_3+\left(3 \
y^2+1\right) I_5\right)-I_3 \left(2 I_5 y^2+\left(y^2+3\right) I_3\right)\right)}{x^2-1}
    \right.\nn\\&&\left.+
\frac{2 x^2 y^4 \left(-I_3^3-I_1 \left(3 I_3+I_5\right) I_3+I_1^2 I_5\right)}{x^2-1}+
\frac{2 x^4 y \left(I_1^2 \left(2 y I_5-2 y^3 I_3\right)-2 y \left(y^2+1\right) I_3^2 I_5\right)}{x^2-1}+
    \right.\nn\\&&\left.
 +\frac{4 x^5 y I_3 \left(2 I_1^2 y^2+I_3 \left(I_5-I_3\right) y^2+I_1 \left(I_3+I_5\right)\right)}{x^2-1}
  \right] \frac{1}{4\pi i}\d_u \log\frac{\Gamma (i u_x-i u_y+1)}{\Gamma (1-i u_x+i u_y)}\nn
\eeqa
where, similarly to $J=2,3$, the integrals go around the branch between $-2g$ and $2g$.

\newpage
\section{Weak coupling expansion -- details}
\label{sec:weakS3}
First, we give the expansion of our results for the slope-to slope functions $\gamma_J^{(2)}$ to 10 loops. We start with $J=2$:
\beqa
\label{weak22long}
	\gamma_{J=2}^{(2)}&=&-8 g^2 \zeta_3+g^4 \left(140 \zeta_5-\frac{32 \pi ^2 \zeta_3}{3}\right)+g^6
   \left(200 \pi ^2 \zeta_5-2016 \zeta_7\right)
	\\ \nn
	&+&g^8 \left(-\frac{16 \pi ^6 \zeta_3}{45}-\frac{88 \pi ^4 \zeta_5}{9}-\frac{9296 \pi ^2 \zeta_7}{3}+27720 \zeta_9\right)
	\\ \nn
	&+&g^{10} \left(\frac{208 \pi ^8 \zeta_3}{405}+\frac{160 \pi ^6 \zeta_5}{27}+144 \pi ^4 \zeta_7+45440 \pi ^2 \zeta_9-377520 \zeta_{11}\right)
	\\ \nn
	&+&g^{12}
   \left(-\frac{7904 \pi ^{10} \zeta_3}{14175}-\frac{17296 \pi ^8 \zeta_5}{4725}-\frac{128 \pi ^6 \zeta_7}{15}-\frac{6312 \pi ^4 \zeta_9}{5}
	\right.
	\\ \nn
	&&\Bigl.\ \ \ \ \ \ \
	-653400 \pi
   ^2 \zeta_{11}+5153148 \zeta_{13}\Bigr)
	\\ \nn
	&+&g^{14} \Bigl(\frac{1504 \pi ^{12} \zeta_3}{2835}+\frac{106576 \pi ^{10} \zeta_5}{42525}-\frac{18992 \pi ^8 \zeta_7}{405}
-\frac{16976 \pi ^6 \zeta_9}{15}
	\Bigr.
	\\ \nn
	&& \Bigl. \ \ \ \ \ \ \
	+\frac{25696 \pi ^4 \zeta_{11}}{9}+\frac{28003976 \pi ^2 \zeta_{13}}{3}-70790720 \zeta_{15}\Bigr)
	\\ \nn
	&+&g^{16}
   \Bigl(-\frac{178112 \pi ^{14} \zeta_3}{382725}-\frac{239488 \pi ^{12} \zeta_5}{127575}+\frac{2604416 \pi ^{10} \zeta_7}{42525}+\frac{8871152 \pi ^8 \zeta_9}{4725}
		\Bigr.
	\\ \nn
	&& \Bigl. \ \ \ \ \ \ \
	+\frac{30157072 \pi ^6 \zeta_{11}}{945}+\frac{8224216 \pi ^4 \zeta_{13}}{45}-133253120 \pi ^2 \zeta_{15}
			\Bigr.
	\\ \nn
	&& \Bigl. \ \ \ \ \ \ \
	+979945824 \zeta_{17}\Bigr)
	\\ \nn
	&+&g^{18}
   \Bigl(\frac{147712 \pi ^{16} \zeta_3}{382725}+\frac{940672 \pi ^{14} \zeta_5}{637875}-\frac{490528 \pi ^{12} \zeta_7}{8505}-\frac{358016 \pi ^{10} \zeta_9}{189}
	\Bigr.
	\\ \nn
	&& \Bigl. \ \ \ \ \ \ \
	-\frac{37441312 \pi ^8 \zeta_{11}}{945}-\frac{9616256 \pi ^6 \zeta_{13}}{15}-\frac{16988608 \pi ^4 \zeta_{15}}{3}
		\Bigr.
	\\ \nn
	&& \Bigl. \ \ \ \ \ \ \
	+1905790848 \pi ^2 \zeta_{17}-13671272160 \zeta_{19}\Bigr)
	\\ \nn
	&+&g^{20} \Bigl(-\frac{135748672 \pi ^{18} \zeta_3}{442047375}-\frac{103683872 \pi ^{16} \zeta_5}{88409475}+\frac{1408423616 \pi
   ^{14} \zeta_7}{29469825}
			\Bigr.
	\\ \nn
	&& \Bigl. \ \ \ \ \ \ \
	+\frac{2288692288 \pi ^{12} \zeta_9}{1403325}+\frac{34713664 \pi ^{10} \zeta_{11}}{945}+\frac{73329568 \pi ^8 \zeta_{13}}{105}
				\Bigr.
	\\ \nn
	&& \Bigl. \ \ \ \ \ \ \
	+\frac{305679296 \pi ^6 \zeta_{15}}{27}+121666688 \pi ^4 \zeta_{17}-27342544320 \pi ^2 \zeta_{19}
					\Bigr.
	\\ \nn
	&& \Bigl. \ \ \ \ \ \ \
	+192157325360 \zeta_{21}\Bigr)
\eeqa

\newpage
Next, for $J=3$,
\beqa
\label{weak23long}
\gamma_{J=3}^{(2)}&=&-2 g^2 \zeta_3+g^4 \left(12 \zeta_5-\frac{4 \pi ^2 \zeta_3}{3}\right)+g^6
   \left(\frac{2 \pi ^4 \zeta_3}{45}+8 \pi ^2 \zeta_5-28 \zeta_7\right)
	\\ \nn
	&+&g^8\;
   \left(-\frac{4 \pi ^6 \zeta_3}{45}-\frac{4 \pi ^4 \zeta_5}{15}-528 \zeta_9\right)
	\\ \nn
	&+&g^{10} \left(\frac{934 \pi ^8 \zeta_3}{14175}+\frac{8 \pi ^6 \zeta_5}{9}-\frac{82 \pi ^4 \zeta_7}{9}-900 \pi ^2 \zeta_9+12870 \zeta_{11}\right)
	\\ \nn
	&+&g^{12} \left(-\frac{572 \pi ^{10} \zeta_3}{14175}-\frac{104 \pi ^8
   \zeta_5}{175}-\frac{256 \pi ^6 \zeta_7}{45}+\frac{2476 \pi ^4 \zeta_9}{9}
	\right. \\ \nn
	&&\ \ \ \ \ \ \ \left.+\frac{57860 \pi ^2 \zeta_{11}}{3}-208208 \zeta_{13}\right)
	\\ \nn
	&+&g^{14}
   \left(\frac{2878 \pi ^{12} \zeta_3}{127575}+\frac{404 \pi ^{10} \zeta_5}{1215}+\frac{326 \pi ^8 \zeta_7}{75}+\frac{3352 \pi ^6 \zeta_9}{135}
	\right. \\ \nn
	&&\ \ \ \ \ \ \ -\left.\frac{80806 \pi ^4 \zeta_{11}}{15}
	-316316 \pi ^2 \zeta_{13}+2994992 \zeta_{15}\right)
	\\ \nn
	&+&g^{16} \left(-\frac{159604 \pi ^{14} \zeta_3}{13395375}-\frac{257204
   \pi ^{12} \zeta_5}{1488375}-\frac{14836 \pi ^{10} \zeta_7}{6075}-\frac{71552 \pi
   ^8 \zeta_9}{2025}
	\right.
	\\ \nn
	&&\ \ \ \ \ \ \  \left.+\frac{4948 \pi ^6 \zeta_{11}}{189}+\frac{4163068 \pi ^4 \zeta_{13}}{45}+\frac{14129024 \pi ^2 \zeta_{15}}{3}-41116608 \zeta_{17}\right)
		\\ \nn
	&+&g^{18}
   \left(\frac{494954 \pi ^{16} \zeta_3}{81860625}+\frac{156368 \pi ^{14} \zeta_5}{1819125}+\frac{6796474 \pi ^{12} \zeta_7}{5457375}+\frac{332 \pi ^{10} \zeta_9}{15}
		\right.
	\\ \nn
	&&\ \ \ \ \ \ \ \left.+\frac{1745318 \pi ^8 \zeta_{11}}{4725} -\frac{868088 \pi ^6 \zeta_{13}}{315}-\frac{22594208 \pi ^4 \zeta_{15}}{15}
	\right. \\ \nn
	&&\ \ \ \ \ \ \ \Bigl.-67084992 \pi ^2 \zeta_{17}+553361016
   \zeta_{19}\Biggr)
	\\ \nn
	&+&g^{20} \left(-\frac{940132 \pi ^{18} \zeta_3}{315748125}-\frac{244456 \pi ^{16} \zeta_5}{5893965}-\frac{29637008 \pi ^{14}
   \zeta_7}{49116375}-\frac{11808196 \pi ^{12} \zeta_9}{1002375}
	\right.
	\\ \nn
		&&\ \ \ \ \ \ \ -\left.\frac{2265364 \pi
   ^{10} \zeta_{11}}{8505}-\frac{68767984 \pi ^8 \zeta_{13}}{14175}+\frac{480208 \pi ^6
   \zeta_{15}}{9}
	\right.
	\\ \nn
		&&\ \ \ \ \ \ \ +\left.
	\frac{71785288 \pi ^4 \zeta_{17}}{3}+934787840 \pi ^2 \zeta_{19}-7390666360 \zeta_{21}\right)
\eeqa

\newpage

Finally, for $J=4$,
\beqa
\label{weak24long}
	\gamma_{J=4}^{(2)}&=&
	g^2 \left(-\frac{14 \zeta_3}{5}+\frac{48 \zeta_5}{\pi ^2}-\frac{252 \zeta_7}{\pi ^4}\right)
	\\ \nn
	&+&g^4
   \left(-\frac{22 \pi ^2 \zeta_3}{25}+\frac{474 \zeta_5}{5}-\frac{8568 \zeta_7}{5 \pi ^2}+\frac{8316
   \zeta_9}{\pi ^4}\right)
\\ \nn
&+&g^6 \left(\frac{32 \pi ^4 \zeta_3}{875}+\frac{3656 \pi ^2 \zeta_5}{175}-\frac{56568 \zeta_7}{25}+\frac{196128 \zeta_9}{5 \pi ^2}-\frac{185328 \zeta_{11}}{\pi
   ^4}\right)
\\ \nn
&+&g^8 \Bigl(-\frac{4 \pi ^6 \zeta_3}{175}-\frac{68 \pi ^4 \zeta_5}{75}-\frac{55312 \pi ^2
   \zeta_7}{125}+\frac{1113396 \zeta_9}{25}-\frac{3763188 \zeta_{11}}{5 \pi ^2}
	\Bigr.
	\\ \nn
	&& \Bigl. \ \ \ \ \ \ \
+\frac{3513510 \zeta_{13}}{\pi ^4}\Bigr)
\\ \nn
&+&g^{10} \Bigl(\frac{176 \pi ^8 \zeta_3}{16875}+\frac{2488 \pi ^6 \zeta_5}{7875}+\frac{2448 \pi ^4 \zeta_7}{125}+\frac{209532 \pi ^2 \zeta_9}{25}-\frac{3969878 \zeta_{11}}{5}
	\Bigr.
	\\ \nn
	&& \Bigl. \ \ \ \ \ \ \
+\frac{13213200 \zeta_{13}}{\pi ^2}-\frac{61261200 \zeta_{15}}{\pi ^4}\Bigr)
\\ \nn
&+&g^{12}
   \Bigl(-\frac{88072 \pi ^{10} \zeta_3}{20671875}-\frac{449816 \pi ^8 \zeta_5}{4134375}-\frac{327212 \pi
   ^6 \zeta_7}{65625}-\frac{338536 \pi ^4 \zeta_9}{875}
	\Bigr.
	\\ \nn
	&& \Bigl. \ \ \ \ \ \ \
-\frac{129520798 \pi ^2 \zeta_{11}}{875}+\frac{66969474 \zeta_{13}}{5}-\frac{220540320 \zeta_{15}}{\pi ^2}
	\Bigr.
	\\ \nn
	&& \Bigl. \ \ \ \ \ \ \
+\frac{1017636048 \zeta_{17}}{\pi ^4}\Bigr)
\\ \nn
&+&g^{14} \Bigl(\frac{795136 \pi ^{12} \zeta_3}{487265625}+\frac{522784 \pi ^{10} \zeta_5}{13921875}+\frac{4021288 \pi ^8 \zeta_7}{2953125}+\frac{1869152 \pi ^6 \zeta_9}{21875}
	\Bigr.
	\\ \nn
	&& \Bigl. \ \ \ \ \ \ \
+\frac{18573952 \pi ^4 \zeta_{11}}{2625}+\frac{62633272 \pi ^2 \zeta_{13}}{25}-\frac{1092799344
   \zeta_{15}}{5}
	\Bigr.
	\\ \nn
	&& \Bigl. \ \ \ \ \ \ \
+\frac{17844607872 \zeta_{17}}{5 \pi ^2}-\frac{16405526592 \zeta_{19}}{\pi ^4}\Bigr)
\\ \nn
&+&g^{16}
   \Bigl(-\frac{30581888 \pi ^{14} \zeta_3}{51162890625}-\frac{43988768 \pi ^{12} \zeta_5}{3410859375}-\frac{446380184 \pi ^{10} \zeta_7}{1136953125}
	\Bigr.
	\\ \nn
	&& \Bigl. \ \ \ \ \ \ \
-\frac{20108936 \pi ^8 \zeta_9}{984375}
-\frac{31755036 \pi ^6 \zeta_{11}}{21875}-\frac{321449336 \pi ^4 \zeta_{13}}{2625}
	\Bigr.
	\\ \nn
	&& \Bigl. \ \ \ \ \ \ \
-\frac{1031925232 \pi ^2 \zeta_{15}}{25}
	+\frac{87296960712 \zeta_{17}}{25}-\frac{283092985656
   \zeta_{19}}{5 \pi ^2}
	\Bigr.
	\\ \nn
	&& \Bigl. \ \ \ \ \ \ \
+\frac{259412389236 \zeta_{21}}{\pi ^4}\Bigr)
\\ \nn
&+&g^{18} \Bigl(\frac{6706432 \pi ^{16}
   \zeta_3}{31672265625}+\frac{816838192 \pi ^{14} \zeta_5}{186232921875}+\frac{2004636572 \pi ^{12} \zeta_7}{17054296875}
	\Bigr.
	\\ \nn
	&& \Bigl. \ \ \ \ \ \ \
+\frac{1950592976 \pi ^{10} \zeta_9}{378984375}
+\frac{2220222512 \pi ^8 \zeta_{11}}{6890625}+\frac{20963856 \pi ^6 \zeta_{13}}{875}
	\Bigr.
	\\ \nn
	&& \Bigl. \ \ \ \ \ \ \
+\frac{254959316 \pi ^4 \zeta_{15}}{125}
+\frac{584553371616 \pi ^2 \zeta_{17}}{875}
	\Bigr.
	\\ \nn
	&& \Bigl. \ \ \ \ \ \ \
-\frac{1375388084412 \zeta_{19}}{25}+\frac{4432313039616 \zeta_{21}}{5 \pi ^2}
-\frac{4049650420200 \zeta_{23}}{\pi ^4}\Bigr)
\eeqa
\beqa
\nn
&+&g^{20}
   \Bigl(-\frac{15308976272 \pi ^{18} \zeta_3}{209512037109375}-\frac{1764947984 \pi ^{16} \zeta_5}{1197211640625}-\frac{18667123736 \pi ^{14} \zeta_7}{517313671875}
	\Bigr.
	\\ \nn
	&& \Bigl. \ \ \ \ \ \ \
-\frac{538293689008 \pi ^{12} \zeta_9}{399070546875}-\frac{657466372 \pi ^{10} \zeta_{11}}{8859375}-\frac{119709052 \pi ^8 \zeta_{13}}{23625}
	\Bigr.
	\\ \nn
	&& \Bigl. \ \ \ \ \ \ \
-\frac{9095498848 \pi ^6 \zeta_{15}}{23625}-\frac{260407748416 \pi ^4 \zeta_{17}}{7875}-\frac{1869110789976 \pi ^2 \zeta_{19}}{175}
	\Bigr.
	\\ \nn
	&& \Bigl. \ \ \ \ \ \ \
+\frac{4293062840352 \zeta_{21}}{5}-\frac{13755955395600 \zeta_{23}}{\pi ^2}+\frac{62673161265000 \zeta_{25}}{\pi
   ^4}\Bigr)
\eeqa

For future reference we have also computed\footnote{As described in the main text (see section 5), in the calculations we used several Mathematica packages for dealing with harmonic sums.} the weak coupling expansion of the anomalous dimensions at order $S^3$, using the known predictions from ABA which are available for any spin at $J=2$ and $J=3$. For $J=2$ we have computed the expansion to three loops\footnote{We remind that in our notation the anomalous dimension is written as $\gamma=\gamma^{(1)}S+\gamma^{(2)}S^2+\gamma^{(3)}S^3+\dots$}:
\beqa
	\gamma_{J=2}^{(3)}&=&g^2\frac{4}{45}\pi^4+g^4 \left(40 \zeta_3^2-\frac{28 \pi ^6}{405}\right)\\ \nn
	&+&g^6 \left(\frac{192}{5} \zeta_{5,3}-\frac{6992 \zeta_3
   \zeta_5}{5}+\frac{280 \pi ^2 \zeta_3^2}{3}+\frac{6962 \pi ^8}{212625}\right)+\mathcal{O}(g^8)
\eeqa
Compared to the $S^2$ part, a new feature is the appearance of multiple zeta values -- here we have $\zeta_{5,3}$, which is defined by
\begin{equation}
	\zeta_{a_1,a_2,\ldots,a_k}=\sum_{0< n_1<n_2<\ldots<n_k<\infty}
	\frac 1{n_{1}^{a_1}n_{2}^{a_2}\ldots n_k^{a_k}}\,
\end{equation}
and cannot be reduced to simple zeta values $\zeta_n$.

For $J=3$ we have obtained the expansion to four loops:
\beqa
\nn
	\gamma_{J=3}^{(3)}&=&\frac{1}{90}\pi^4g^2+g^4\(4 \zeta_3^2+\frac{\pi ^6}{1890}\)
	+g^6\(4\zeta_{5,3}+4 \pi ^2 \zeta_3^2-72 \zeta_3 \zeta_5-\frac{2
   \pi ^8}{675}\)\\ \nn
   &+&g^8\left(-112 \zeta_{2,8}+\frac{20}{3} \pi ^2 \zeta_{5,3}+728 \zeta_3
   \zeta_7+448 \zeta_5^2-\frac{224}{3} \pi ^2 \zeta_3
   \zeta_5\right.\\
	&& \left.+\frac{4 \pi ^4 \zeta_3^2}{5}-\frac{41 \pi
   ^{10}}{133650}\right)
   +\mathcal{O}(g^{10})
\eeqa

\section{Higher mode numbers}
\label{sec:appHigher}
\subsection{Slope function for generic filling fractions and mode numbers}
\label{sec:Sanyn}
Let us extend the discussion of section \ref{sec:evenLsol} by considering the state corresponding to a solution of the asymptotic Bethe equations with arbitrary mode numbers and filling fractions\footnote{For simplicity we also consider even $J$ here.}.
We expect that in the $\P\mu$ system this should correspond to\footnote{We no longer expect $\mu_{24}$ to be either even or odd, since in the Bethe ansatz description of the state with generic mode numbers and filling fractions the Bethe roots are not distributed symmetrically.}
\beq
	\mu_{24} = \sum_{n=-\infty}^\infty C_n e^{2\pi n u}.
\eeq
As an example, for the ground state twist operator we have $\mu_{24}=\sinh(2\pi u)$, which is reproduced by choosing $C_{-1} = -1/2, C_1 = 1/2$ and all other $C$'s set to $0$.

It is straightforward to solve the $\P\mu$ system in the same way as in section \ref{sec:evenLsol}, and we find the energy
\beq
	\gamma = \frac{\sqrt{\lambda}}{J} \frac{\sum_n C_n I_{J+1}(n \sqrt{\lambda}) }{\sum_n C_n I_{J}(n \sqrt{\lambda})/n} \, S,
\eeq
which can also be written in a more familiar form as
\beq
	\gamma = \sum_n \alpha_n \frac{n\sqrt{\lambda}}{J} \frac{I_{J+1}(n \sqrt{\lambda})}{I_{J}(n \sqrt{\lambda})} \, S,
\eeq
where
\beq
	\alpha_n = \frac{C_n I_{J}(n \sqrt{\lambda}) / n}{\sum_m C_m I_{J}(m \sqrt{\lambda}) / m}
\eeq
are the filling fractions.

The coefficients $C_n$ are additionally constrained by
\beq
	\sum_n C_n I_J(n\sqrt{\lambda}) = 0,
\eeq
which ensures that the $\bP_a$ functions have correct asymptotics. This constraint implies a relation between the filling fractions,
\beq
	\sum_n n \, \alpha_n = 0,
\eeq
which is also familiar from the asymptotic Bethe ansatz.

\subsection{Curvature function and higher mode numbers}
\label{sec:appN}

In the main text we discussed the NLO solutions to the $\bP\mu$ system which are based on the leading order solutions \eqref{eq:P1solLOevenL}-\eqref{eq:P4solLOevenL} or \eqref{P1oddL}-\eqref{P4oddL}. One of the assumptions for constructing the leading order solution was to allow $\mu_{ab}$ to have only $e^{\pm 2\pi u}$ in asymptotics at infinity (we recall that this led to all $\mu$'s being constatnt except $\mu_{24}$ which is equal to $\sinh\({2\pi u}\)$ or $\cosh\({2\pi u}\)$), while in principle requiring $\mu_{ab}$ to be periodic one could also allow  to have $e^{2n\pi u}$ with any integer $n$.
Thus a natural generalization of the leading order solution is to consider $\mu_{24}=\sinh\({2\pi n u}\)$ or $\mu_{24}=
\cosh\({2\pi n u}\)$, where $n$ is an arbitrary integer. As discussed above (see the end of section \ref{sec:LOresultevenL} and appendix \ref{sec:Sanyn}), we believe that at the leading order in $S$ such solutions correspond to states with mode numbers equal to $n$, and they reproduce the slope function for this case.

Proceeding to order $S^2$, the calculation of the curvature function $\gamma^{(2)}(g)$ with $\mu_{24}=\sinh\({2\pi n u}\)$ or $\mu_{24}=
\cosh\({2\pi n u}\)$ can be done following the same steps as for $n=1$.
The final results for $J=2,3$ and $4$ are given by exactly the same formulas as for $n=1$ (\eqref{gamma2L2}, \eqref{gamma2L3} and \eqref{gamma2L4} respectively) -- the only difference is that now one should set in those expressions
\beqa
\label{nrep1}
&&I_k=I_k(4\pi n g ),\\
&&\sh_-^x=\[\sinh\({2 \pi n u_x}\)\]_-, \\
&&\sh_-^y=\[\sinh\({2 \pi n u_y}\)\]_-, \\
&&\ch_-^x=\[\cosh\({2 \pi n u_x}\)\]_-, \\
\label{nreplast}
&&\ch_-^y=\[\cosh\({2 \pi n u_y}\)\]_-.
\eeqa
It would be natural to assume that this solution of the $\bP\mu$ system describes anomalous dimensions for states with mode number $n$ at order $S^2$. However we found some peculiarities in the strong coupling expansion of the result. The strong coupling data available for comparison in the literature for states with $n>1$ also relies on some conjectures (see \cite{Basso:2011rs}, \cite{Gromov:2011bz}), so the interpretation of this solution is not fully clear to us.

The weak coupling expansion for this case turns out to be related in a simple way to the $n=1$ case. One should just replace $\pi\to n\pi$ in the expansions for $n=1$ which were given in \eq{weak22long}, \eq{weak23long}, \eq{weak24long}. For example,
\beqa
\nn
	\gamma_{J=2}^{(2)}&=&-8 g^2 \zeta_3+g^4 \left(140 \zeta_5-\frac{32 n^2\pi ^2 \zeta_3}{3}\right)+g^6
   \left(200 n^2\pi ^2 \zeta_5-2016 \zeta_7\right)
	+\dots
	\\
\eeqa
It would be interesting to compare these weak coupling predictions to results obtained from the asymptotic Bethe ansatz (or by other means) as it was done for $n=1$ in section \ref{sec:weak}.

Let us now discuss the strong coupling expansion. According to Basso's conjecture \cite{Basso:2011rs} (see also \cite{Gromov:2011bz}), the structure of the expansion may be obtained from
\beq
\label{basso1app}
\Delta^2=J^2+S
\(
A_1\sqrt{\mu}+A_2+\dots
\)
+S^2
\(
B_1+\frac{B_2}{\sqrt\mu}
+\dots
\)
+S^3
\(
\frac{C_1}{\mu^{1/2}}
+\frac{C_2}{\mu^{3/2}}
+\dots
\)
+{\cal O}({ S}^4)\;,
\eeq
where $\mu=n^2\lambda$. This gives
\beqa
\label{basso2app}
\Delta&=&J+\frac{S}{2J}
\(
A_1n\sqrt{\lambda}+A_2+\frac{A_3}{n\sqrt{\lambda}}+\dots
\)\\
\nn&+&S^2
\(
- \frac{A_1^2}{8J^3} \, n^2\lambda
-  \frac{A_1A_2}{4J^3} \, n\sqrt{\lambda}
+\[\frac{B_1}{2J}-\frac{A_2^2+2A_1 A_3}{8J^3}\]
+
\[
\frac{B_2}{2J}
-\frac{A_2A_3+A_1A_4}{4J^3}
\]  \frac{1}{n\sqrt\lambda}
+\dots
\)
+{\cal O}(S^3)\;.
\eeqa
where $A_i$ are known from Basso's slope function. Substituting them, we find
\beq
\label{gamma2basso}
	\gamma^{(2)}_{J}(g)=-\frac{8 \pi ^2g^2 n^2}{J^3}+\frac{2 \pi  g n}{J^3}+\frac{B_1-1}{2 J}+\frac{8 B_2 J^2-4 J^2+1}{64 \pi  g J^3 n}+\dots
\eeq
However, already in \cite{Gromov:2011bz} some inconsistencies were found if one assumes this structure for $n>1$. Let us extend that analysis by comparing the prediction \eq{gamma2basso} to our results from the $\bP\mu$-system. To compute the expansion of our results, similarly to the $n=1$ case, we evaluated $\gamma_{J}^{(2)}(g)$ numerically for many values of $g$, and then fitted the result by powers of $g$. As for $n=1$ we found with high precision (about $\pm10^{-16}$) that the first several coefficients involve only rational numbers and powers of $\pi$. Our results for $n=2,3$ and $J=2,3,4$ are summarized below:
\beq
\label{gamma2L22}
	\gamma^{(2)}_{J=2,n=2}(g)=-4\pi^2g^2+\frac{\pi g}{2}+\frac{17}{8}-\frac{0.29584877037648771(2)}{g}
	+\dots
\eeq
\beq
	\gamma^{(2)}_{J=3,n=2}(g)=-\frac{32}{27} \pi ^2 g^2+\frac{4 \pi  g}{27}+\frac{17}{12}
	-\frac{0.2928304112866493(9)}{g}
	+\dots
\eeq
\beq
\gamma^{(2)}_{J=4,n=2}(g)=-\frac{1}{2} \pi ^2 g^2+\frac{\pi  g}{16}+\frac{17}{16}-\frac{0.319909936615448(9)}{g}
+\dots
\eeq
\beq
	\gamma^{(2)}_{J=2,n=3}(g)=-9 \pi ^2 g^2+\frac{3 \pi  g}{4}+\frac{23}{4}-\frac{0.8137483(9)}{g}+\dots
\eeq
\beq
	\gamma^{(2)}_{J=3,n=3}(g)=-\frac{8}{3} \pi ^2 g^2+\frac{2 \pi  g}{9}+\frac{23}{6}-\frac{0.892016609(2)}{g}
	+\dots
\eeq
\beq
	\gamma^{(2)}_{J=4,n=3}(g)=-\frac{9}{8} \pi ^2 g^2+\frac{3 \pi  g}{32}+\frac{23}{8}-\frac{1.035945580(6)}{g}+\dots
\eeq
Here in the coefficient of $\ofrac{g}$ the digit in brackets is the last known one within our precision\footnote{We did not seek to achieve high precision in this coefficient for $n=3$.}.

Comparing to \eq{gamma2basso} we find full agreement in the first two terms (of order $g^2$ and of order $g$). The next term in \eq{gamma2basso} (of order $g^0$) is determined by $B_1$, which in \cite{Gromov:2011bz} was found to be
\beq
	B_1=\frac{3}{2}
\eeq
for all $n,J$, based on consistency with the classical energy. However, comparing our results with \eq{gamma2basso} we find a different value:
\beqa
	B_1&=&\frac{19}{2} \ \  \text{for }\ n=2\;,\\ \nn
	B_1&=&23 \ \ \, \text{for }\ n=3\;.
\eeqa
For both $n=2$ and $n=3$ this prediction for $B_1$ is independent of $J$.

The next term is of order $\ofrac{g}$ and is determined by $B_2$, which in \cite{Gromov:2011bz} was fixed to
\beq
B_2=
\left\{
\bea{ll}
-3\,\zeta_3+\frac{3}{8}&\;\;,\;\;n=1\\
-24\,\zeta_3-\frac{13}{8}&\;\;,\;\;n=2\\
-81\,\zeta_3-\frac{24}{8}&\;\;,\;\;n=3
\eea
\right.\;.
\eeq
However, this does not agree with our numerical predictions for $n=2$ and $3$. Furthermore, for $n=2$ we extracted the coefficient of $\ofrac{g}$ with high precision (about $10^{-17}$, see \eq{gamma2L22}) but were unable to fit it as a combination of simple zeta values using the EZ-Face calculator \cite{ezface}.

Thus our results appear to disagree with the values of $B_1$ and $B_2$ obtained in \cite{Gromov:2011bz}, but how to interpret this is not clear to us. Although our solution of the $\bP\mu$-system for $n>1$ looks fine at order $S$, it may be that to capture anomalous dimensions at order $S^2$ some other solution should be used.
Another option is that the ansatz for the structure of anomalous dimensions at strong coupling may need to be modified when $n>1$ (as already suspected in \cite{Gromov:2011bz}), and our results may help provide some guidance in this case.

\end{document}